\documentclass[sigconf,nonacm]{acmart}

\usepackage{subfigure}
\usepackage{siunitx}
\usepackage{enumitem}

\usepackage{algorithm}
\usepackage[noend]{algpseudocode}
\let\ReturnInline\Return
\renewcommand{\Return}{\State\ReturnInline}
\algrenewcommand\algorithmicrequire{$\rhd$}
\algrenewcommand\algorithmicensure{$\square$}

\AtBeginDocument{%
  \providecommand\BibTeX{{%
    \normalfont B\kern-0.5em{\scshape i\kern-0.25em b}\kern-0.8em\TeX}}}

\setcopyright{acmcopyright}
\copyrightyear{2018}
\acmYear{2018}
\acmDOI{XXXXXXX.XXXXXXX}

\acmConference[Conference acronym 'XX]{Make sure to enter the correct
  conference title from your rights confirmation emai}{June 03--05,
  2018}{Woodstock, NY}

\newcommand{\ignore}[1]{}

\begin{document}

\title[Performance Comparison of Graph Representations Which Support Dynamic Graph Updates]{Performance Comparison of Graph Representations \\Which Support Dynamic Graph Updates}


\author{Subhajit Sahu}
\email{subhajit.sahu@research.iiit.ac.in}
\affiliation{%
  \institution{IIIT Hyderabad}
  \streetaddress{Professor CR Rao Rd, Gachibowli}
  \city{Hyderabad}
  \state{Telangana}
  \country{India}
  \postcode{500032}
}


\settopmatter{printfolios=true}

\begin{abstract}
Research in graph-structured data has grown rapidly due to graphs' ability to represent complex real-world information and capture intricate relationships, particularly as many real-world graphs evolve dynamically through edge/vertex insertions and deletions. This has spurred interest in programming frameworks for managing, maintaining, and processing such dynamic graphs. In this report, we evaluate the performance of PetGraph (Rust), Stanford Network Analysis Platform (SNAP), SuiteSparse:GraphBLAS, cuGraph, Aspen, and our custom implementation in tasks including loading graphs from disk to memory, cloning loaded graphs, applying in-place edge deletions/insertions, and performing a simple iterative graph traversal algorithm. Our implementation demonstrates significant performance improvements: it outperforms PetGraph, SNAP, SuiteSparse:GraphBLAS, cuGraph, and Aspen by factors of $177\times$, $106\times$, $76\times$, $17\times$, and $3.3\times$ in graph loading; $20\times$, $235\times$, $0.24\times$, $1.3\times$, and $0\times$ in graph cloning; $141\times$/$45\times$, $44\times$/$25\times$, $13\times$/$11\times$, $28\times$/$34\times$, and $3.5\times$/$2.2\times$ in edge deletions/insertions; and $67\times$/$63\times$, $86\times$/$86\times$, $2.5\times$/$2.6\times$, $0.25\times$/$0.24\times$, and $1.3\times$/$1.3\times$ in traversal on updated graphs with deletions/insertions.
\end{abstract}

\keywords{Graph representation, Dynamic graph updates}


\maketitle

\section{Introduction}
\label{sec:introduction}
Graph-structured data has become a cornerstone of modern data analysis, enabling the representation of complex relationships in domains such as social networks, biological systems, and recommendation engines. The dynamic nature of many real-world graphs --- where edges and vertices are frequently added or removed --- has further intensified the need for efficient frameworks capable of managing and processing such dynamic graphs.

\begin{figure}[hbtp]
  \centering
  \subfigure{
    \label{fig:vector2d-runtime--all}
    \includegraphics[width=0.98\linewidth]{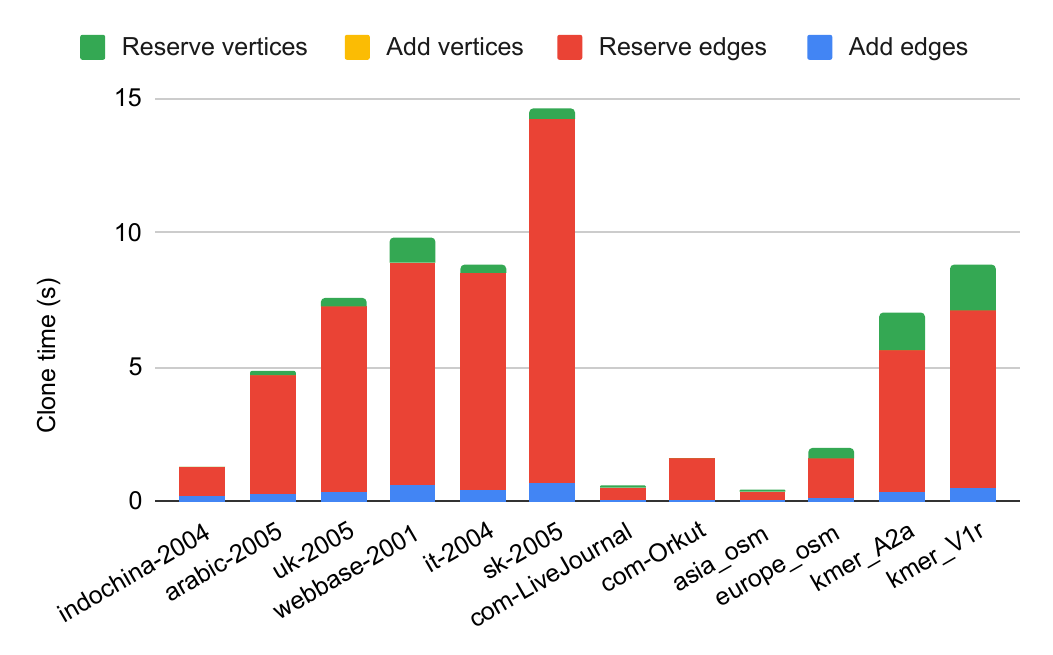}
  } \\[-2ex]
  \caption{Runtime split, in seconds, for cloning a graph based on a \textit{vector2d} representation is shown for each graph in Table \ref{tab:dataset}. As illustrated, an average of $74\%$ of the total runtime is spent on memory allocation for the edges.}
  \label{fig:vector2d-runtime}
\end{figure}

However, despite significant advancements in graph processing frameworks, there remains a gap in performance, especially when handling large-scale, dynamic graphs. Existing frameworks, such as PetGraph, Stanford Network Analysis Platform (SNAP), SuiteSparse:GraphBLAS, cuGraph, and Aspen, offer varying degrees of efficiency but often struggle with scalability and speed when dealing with massive datasets or frequent updates. One of the primary bottlenecks is memory allocation during dynamic operations, particularly in graph cloning, where a significant portion of runtime is spent on edge memory allocation. As illustrated in Figure \ref{fig:vector2d-runtime}, an average of $74\%$ of the total runtime for cloning a graph based on a \textit{vector2d} representation is consumed by memory allocation for edges. This inefficiency highlights the need for optimized graph representations that minimize memory allocation overhead.

In this technical report, we present a comprehensive evaluation of several state-of-the-art graph processing frameworks, including our \textbf{DiGraph} implementation,\footnote{\url{https://github.com/puzzlef/graph-openmp}} which leverages our Concurrent Power-of-2 Arena Allocator (CP2AA)\footnote{\url{https://github.com/puzzlef/allocator-openmp}} for efficient memory management. We compare the performance of these frameworks across four key tasks: \textbf{(1)} Loading graphs from disk into memory, \textbf{(2)} Cloning graphs, \textbf{(3)} Applying in-place edge insertions and deletions, and \textbf{(4)} Executing a simple iterative graph traversal algorithm.

Our results demonstrate that our DiGraph benefits from several optimizations: Algorithm \ref{alg:load} significantly improves graph loading times, Algorithm \ref{alg:clone} enables efficient deep copies, and Algorithms \ref{alg:sub} and \ref{alg:add} facilitate efficient in-place batch updates. Additionally, the use of contiguous edge arrays and a Struct-of-Arrays (SoA) approach enhances the execution of graph algorithms, such as $k$-step reverse walks. We also observe that SuiteSparse:GraphBLAS's lazy copying and, in particular, Aspen's zero-cost snapshotting can significantly improve the performance of applying dynamic batch updates to the given graph when creating new graph instances.
\ignore{Finally, our DiGraph could further benefit from optimizations for small batch updates.}

This technical report is organized as follows: Section \ref{sec:related} reviews related work, providing an overview of existing state-of-the-art graph processing frameworks. Section \ref{sec:approach} describes our custom graph representation and algorithms in detail. Section \ref{sec:evaluation} presents our experimental setup and discusses the performance results. Finally, Section \ref{sec:conclusion} presents some concluding remarks and future directions.

\section{Related work}
\label{sec:related}
PetGraph \cite{sverdrup2025petgraph} is a general-purpose graph library written in Rust. It offers several graph implementations, each with its own tradeoffs. We find its \texttt{GraphMap} structure to offer the best performance for the workload we are testing, and hence we focus on it in this report. A \texttt{GraphMap} uses an associative array (i.e. a hashmap) to represent vertices, and an inner hashmap for each vertex to represent its neighbors. This allows it to achieve quick lookup ($O(1)$ average per neighbor check), and testing whether an edge exists is simply a matter of doing two hashmap lookups. PetGraph provides support for single-edge operations, and these can be invoked in a loop for batch updates. When an edge is added (using \texttt{add\_edge()}), PetGraph first locates the source node in its hashmap and then inserts the new destination ID into the node's inner hashmap. Similarly, deleting an edge (using \texttt{remove\_edge()}) involves locating the appropriate neighbor entry in the inner hashmap and then removing it.

Stanford Network Analysis Platform (SNAP) \cite{leskovec2016snap} is an open-source C++ library (with a Python interface, Snap.py) for analyzing and manipulating large-scale networks. It supports undirected graphs (TUNGraph), directed graphs (TNGraph), and multigraphs (TNEGraph). Nodes are stored in hash tables, keyed by unique integer IDs. Each node maintains one (for undirected graphs) or two sorted vectors (outgoing and incoming edges, for directed graphs) listing its neighbors. Sorting these vectors facilitates fast lookup via binary search while keeping the memory footprint compact. SNAP builds on top of vectors and hash tables (similar to the STL but optimized for SNAP's use cases) that are implemented in the GLib library developed at the Jozef Stefan Institute. It also provides support for single-edge operations, and these can be invoked in a loop for batch updates. When an edge is added, SNAP first locates the source node in its hash table ($O(1)$ average time) and then inserts the new destination ID into the node's sorted neighbor vector. Similarly, deleting an edge involves locating the appropriate neighbor entry in the sorted vector and then removing it.

SuiteSparse:GraphBLAS \cite{davis2023algorithm, davis2019algorithm} is an open-source C library providing an efficient parallel implementation of the GraphBLAS standard for (sparse) matrix-based graph algorithms. GraphBLAS objects (matrices and vectors) are opaque and stored in standard compressed formats: Compressed Sparse Row (CSR) or Compressed Sparse Column (CSC). In these formats, a matrix comprises \textbf{(1)} a pointer array indicating the start of each column/row, \textbf{(2)} an index array storing row indices (CSC) or column indices (CSR), and \textbf{(3)} a value array holding nonzero entries. Dynamic graph updates are handled via \textit{zombies} and \textit{pending tuples}: deleted edges are marked as zombies by modifying indices instead of immediate removal, preventing costly restructuring, while new edge insertions are temporarily stored as pending tuples in an unsorted list, enabling fast incremental updates in $O(\log k)$ time (for $k$ entries in a column). A subsequent assembly phase consolidates these pending updates by removing zombie entries, and flushing pending tuples by sorting and merging them into the data arrays. This lazy consolidation amortizes the cost of incremental updates\ignore{over multiple operations}, and is only done when an operation is invoked that requires the matrix to be in a fully assembled state.

cuGraph \cite{kang2023cugraph} is NVIDIA's GPU-accelerated graph analytics library that is part of the RAPIDS suite. cuGraph accepts graph data in various forms --- edge lists or adjacency lists --- and can ingest data stored in GPU DataFrames provided by cuDF. This makes it easy to integrate with RAPIDS workflows, where data preparation and graph processing can be done entirely on the GPU. Internally, cuGraph represents graphs in a CSR-like format, where row offsets mark the start of each vertex's edge list, column indices store neighbor vertex IDs, and optional edge attributes hold weights for weighted graphs. Building the CSR representation typically requires sorting the edges, performing parallel scan to compute the row offsets, and copying the neighbor indices. Performing a batch update of edge insertions involves a parallel merge with the existing sorted edge list and removing duplicates before creating a new graph. For deletions, edges marked for removal are filtered out and a new graph is built from the remaining edges.

Aspen \cite{dhulipala2019low} is a low-latency graph streaming framework designed to support concurrent graph queries and dynamic updates on large graphs, and has been shown to outperform other dynamic graph streaming systems (e.g., Stinger and LLAMA) in both throughput and memory usage. Its core idea is to represent the graph using compressed purely-functional search trees, called C-trees. Both the vertex set and the adjacency lists are represented using C-trees. The vertex set is maintained in a persistent (immutable) vertex-tree, and each vertex's adjacency list is stored in an edge-tree that is itself a C-tree. In a C-tree, instead of storing one element per node, elements are ``chunked" into arrays. This not only reduces the number of nodes (thus saving memory) but also improves cache locality. For integer data (typical for vertex IDs and edges), Aspen applies difference encoding within these chunks to further compress the representation. Because the C-trees are purely functional (immutable) \cite{okasaki1998purely}, every update creates a new version by copying only the nodes along the modified path (copy-on-write). This design enables lightweight snapshots and concurrent operations: readers can acquire a consistent snapshot of the graph instantly (this is simply a pointer, or handle, to the graph's root), while a single writer can apply updates without blocking queries.

We now describe how Aspen loads a graph, clones it to create snapshots, and applies batch updates of edge insertions and deletions. Aspen expects input graphs in a variant of the “adjacency graph” format (similar to what Ligra uses). The file typically starts with metadata (e.g., number of vertices and edges) followed by offsets and edge lists. The loader reads the graph file and constructs a persistent vertex-tree (using a framework such as PAM for augmented maps). Each vertex is inserted along with its metadata. For each vertex, the corresponding list of out-edges is read and converted into an edge-tree (a C-tree). This process involves “chunking” the sorted edge list and applying difference encoding for compression. For applying batch updates of edge insertions/deletions, Apsen provides functions such as \texttt{insert\_edges\_batch()} and \texttt{delete\_edges\_batch()}. Both functions first preprocess the batch of updates to group them by source vertex and aggregate the new edges (or edges to delete) for each vertex. Then, for each source vertex, the corresponding edge-tree is updated using a \textit{MultiInsert} or \textit{MultiDelete} operation. These operations are efficient because they operate on chunks and only require updating the nodes along the modified paths. While only one writer is allowed to update at a time, the batch processing is designed to be highly parallel internally, allowing different vertex updates to be processed concurrently in the union/difference steps if they affect different parts of the tree. Aspen's API also includes functions to manage graph versions and handle garbage collection (reclaiming old versions when unused).

Finally, we discuss how Aspen handles memory management. Aspen employs a parallel reference counting garbage collector and a custom pool-based allocator to manage memory efficiently. It uses a concurrent memory allocator, \texttt{list\_allocator}, which minimizes contention by maintaining per-thread local memory pools. Each thread primarily operates on its local pool, fetching batches from a global stack as needed and returning excess memory when the pool surpasses a threshold. Memory is structured in a linked-list format and aligned to cache line boundaries to prevent false sharing.

\section{Approach}
\label{sec:approach}
\subsection{Our Graph Representation}
\label{sec:digraph}

Our directed graph implementation, \textbf{DiGraph}, which leverages our Concurrent Power-of-2 Arena Allocator (CP2AA) for efficient memory management (see Section \ref{sec:cp2aa} for details), is described in Algorithms \ref{alg:digraph1} and \ref{alg:digraph2}. The implementation supports operations for inserting and deleting a batch of edges, where a batch of edges is represented using \texttt{DiGraph}. Thus, insertion of a batch of edges corresponds to a graph union operation, while deletion of a batch of edges corresponds to a graph subtraction operation.

The initialization of a \texttt{DiGraph} (lines \ref{alg:digraph--struct-begin}-\ref{alg:digraph--struct-end}) involves setting up four main data structures: \textbf{(1)} A bit array, $exists$, to track vertex existence; \textbf{(2)} An adjacency list, $edges$, to store outgoing edges; \textbf{(3)} An array, $degrees$, to track each vertex's out-degree; and \textbf{(4)} An array, $capacities$, to manage allocated edge storage per vertex. Memory allocation is handled by the CP2AA allocator, $cp2aa$. Additionally, the initialization defines key variables: reserved memory ($RES$), vertex capacity ($CAP$, the maximum vertex ID + 1), total vertices ($N$), and total edges ($M$). Configured constants include $\textsc{bool\_bits}$, which inidicates the granularity of vertex existence flags, and $\textsc{edge\_size}$, which represents the size of an edge in bytes.

The function \texttt{hasVertex()} (lines \ref{alg:digraph--has-vertex-begin}-\ref{alg:digraph--has-vertex-end}) of \texttt{DiGraph} checks whe-ther a given vertex exists in the graph by verifying if its ID falls within the allocated range ($CAP$) and whether the corresponding bit in $exists$ is set. The function \texttt{degree()} (lines \ref{alg:digraph--degree-begin}-\ref{alg:digraph--degree-end}) returns the number of outgoing edges for a vertex, defaulting to zero if the vertex is not within range. The \texttt{edges()} function (lines \ref{alg:digraph--edges-begin}-\ref{alg:digraph--edges-end}) retrieves the outgoing edges of a vertex, returning an empty set if the vertex does not exist.
The \texttt{reserve()} function (lines \ref{alg:digraph--reserve-begin}-\ref{alg:digraph--reserve-end}) ensures that sufficient space is allocated to accommodate new vertices. It first calculates a new reserved size by rounding up to the system's page size. If the requested capacity is greater than the current vertex capacity $CAP$, memory is reallocated (in parallel) using the \texttt{reallocate()} function (lines \ref{alg:digraph--reallocate-begin}-\ref{alg:digraph--reallocate-end}), which adjusts the size of multiple arrays, including $exists$ (bit array), $edges$, $degrees$, and $capacities$. Finally, the the graph’s vertex capacity $CAP$ and the reserved memory size $RES$ are updated.
The \texttt{reallocate()} function (lines \ref{alg:digraph--reallocate-begin}–\ref{alg:digraph--reallocate-end}) is responsible for resizing memory allocations while preserving existing data where possible. It first checks if the requested reserved amount $R\_1$ matches the current reserved amount $R\_0$. If they are equal, no new allocation is needed, and only the newly added portion of the array (from $N\_0$ to $N\_1$) is initialized to zero in parallel. Otherwise, new memory is allocated for $R\_1$ entries, and the existing data is copied up to the minimum of the old and new sizes ($M = min(N\_0, N\_1)$). Any remaining entries in the newly allocated space (from $M$ to $N\_1$) are initialized to zero in parallel. Once the data transfer is complete, the old memory is deallocated, and the newly allocated pointer is returned.

The \texttt{addVertex()} function (lines \ref{alg:digraph--add-vertex-begin}-\ref{alg:digraph--add-vertex-end}) ensures that a vertex exists by expanding storage if necessary and then setting its existence flag. 
The function \texttt{allocateEdges()} (lines \ref{alg:digraph--allocate-edges-begin}-\ref{alg:digraph--allocate-edges-end}) is responsible for allocating memory for a vertex’s outgoing edges using the CP2AA allocator. It first checks whether the vertex ID $u$ exceeds the current capacity ($CAP$) or if memory for its edges has already been allocated. If either condition is met, the function returns immediately. Otherwise, it determines the required memory size in bytes by multiplying the desired number of edges ($deg$) by $\textsc{edge\_size}$, which represents the storage size of an edge (a tuple of a destination vertex and an edge weight). The computed memory requirement is then passed to the CP2AA allocator's \texttt{allocationSize()} function, which adjusts the requested size to make it a power-of-2 size (in bytes), or a multiple of system page size. The function then allocates the necessary amount of memory (in bytes) using $cp2aa.allocate()$, stores the resulting pointer in $edges[u]$, and updates $capacities[u]$ to reflect the allocated number of edges.

Adding an edge without additional safety checks is handled by \texttt{addEdgeUnsafe()} (lines \ref{alg:digraph--add-edge-unsafe-begin}-\ref{alg:digraph--add-edge-unsafe-end}), which retrieves the pointer to the edge list $edges[u]$, increments the degree counter $degrees[u]$ atomically, and inserts the new edge.
The \texttt{addEdges()} function (lines \ref{alg:digraph--add-edges-begin}-\ref{alg:digraph--add-edges-end}) is responsible for adding a list of outgoing edges to a vertex while ensuring proper memory allocation. If the vertex does not exist or the provided edge list is empty, the function returns immediately. First, it records the current degree of the vertex ($deg\_prev$) and calculates the maximum possible degree after adding the new edges ($deg\_max$). It then determines the required memory allocation size using the CP2AA allocator and allocates a new memory block for storing the updated edge list. The function merges the existing edges with the new edges using the \texttt{setUnion()} operation, ensuring they are stored in the allocated memory. Note that edges in the list must be sorted. Once the new edges are successfully added, the function deallocates the previously allocated memory using CP2AA, updates the edge pointer, and records the new edge capacity. Finally, it returns the number of newly added edges by computing the difference between the updated and previous degrees.
The \texttt{removeEdges()} function (lines \ref{alg:digraph--remove-edges-begin}-\ref{alg:digraph--remove-edges-end}) is responsible for removing a specified list of outgoing edges from a given vertex. First, it checks whether the vertex exists in the graph using \texttt{hasVertex()} and whether the provided edge list is non-empty. If either condition is not met, the function returns immediately with a value of zero. Otherwise, it records the current out-degree of the vertex in $deg\_prev$. The function then performs a set difference operation to remove the specified edges from the vertex’s adjacency list, effectively updating the edge count. Finally, it returns the number of edges removed, calculated as the difference between the original degree and the updated degree of the vertex.

Finally, the \texttt{update()} function (lines \ref{alg:digraph--update-begin}-\ref{alg:digraph--update-end}) ensures the integrity of the graph by sorting and deduplicating edges while updating the total vertex and edge counts. First, if the edges are not already sorted, each vertex's edge list is sorted in parallel using the \texttt{sortByKey()} function. Next, if duplicate edges exist, they are removed via \texttt{uniqueByKey()}, also executed in parallel. After enforcing order and uniqueness, the function recalculates the total number of vertices, $N$, and edges, $M$, by iterating over all possible vertex IDs up to the current capacity, $CAP$.\ignore{If a vertex is marked as existing, its count is incremented, and its degree contributes to the total edge count.}

\begin{algorithm}[hbtp]
\caption{Our Directed Graph that uses CP2AA allocator.}
\label{alg:digraph1}
\begin{algorithmic}[1]
\Require{$T_K, T_W$: Datatype for vertex ID, edge weight}
\Require{$\textsc{pool\_size}_a$: Size of each memory pool (constant)}

\Statex

\State \textbf{struct} $DiGraph \langle T_K, T_W \rangle ()$ \label{alg:digraph--struct-begin}
\State \ \ $\textsc{bool\_bits} \gets 64$ \Comment{Boolean stored in 64-bit chunks}
\State \ \ $\textsc{edge\_size} \gets sizeof((T_K, T_W))$ \Comment{Size of an edge in bytes}
\State \ \ $exists \gets \{\}$ \Comment{Vertex existence flags}
\State \ \ $edges \gets \{\}$ \Comment{Outgoing edges for each vertex}
\State \ \ $degrees \gets \{\}$ \Comment{Out-degree of each vertex}
\State \ \ $capacities \gets \{\}$ \Comment{Edge capacity of each vertex}
\State \ \ $cp2aa \gets CP2AA \langle \textsc{pool\_size}_a \rangle ()$ \Comment{Memory allocator}
\State \ \ $CAP \gets 0$ \Comment{Vertex capacity (max vertex id + 1)}
\State \ \ $RES \gets 0$ \Comment{Memory reserved for vertices}
\State \ \ $N \gets 0$ \Comment{Total number of vertices}
\State \ \ $M \gets 0$ \Comment{Total number of edges} \label{alg:digraph--struct-end}

\Statex

\State $\rhd$ Check if a vertex exists in the graph
\Function{hasVertex}{$u$} \textbf{of} DiGraph \label{alg:digraph--has-vertex-begin}
  \Return{$u < CAP$ \textbf{and} $getBit(exists, u)$}
\EndFunction \label{alg:digraph--has-vertex-end}

\Statex

\State $\rhd$ Get the number of outgoing edges of a vertex
\Function{degree}{$u$} \textbf{of} DiGraph \label{alg:digraph--degree-begin}
  \Return{$u < CAP$? $degrees[u]$ : $0$}
\EndFunction \label{alg:digraph--degree-end}

\Statex

\State $\rhd$ Get the outgoing edges of a vertex
\Function{edges}{$u$} \textbf{of} DiGraph \label{alg:digraph--edges-begin}
  \Return{$u < CAP$? $edges[u][0 \dots degrees[u]]$ : \{\}}
\EndFunction \label{alg:digraph--edges-end}

\Statex

\State $\rhd$ Reserve space for a number of vertices
\Function{reserve}{$n$} \textbf{of} DiGraph \label{alg:digraph--reserve-begin}
  \State $n \gets max(n, CAP)$
  \State $\rhd$ Compute new reserved size (round up to page size)
  \State $res \gets \lceil n / \textsc{page\_size} \rceil * \textsc{page\_size}$
  \If{$n \leq CAP$ \textbf{and} $res = RES$} \ReturnInline{}
  \EndIf
  \State $\rhd$ Allocate new memory
  \State $B \gets \textsc{bool\_bits}$
  \State $exists \gets reallocate(exists, \lceil \frac{CAP}{B} \rceil, \lceil \frac{RES}{B} \rceil, \lceil \frac{n}{B} \rceil, \lceil \frac{res}{B} \rceil)$
  \State $edges \gets reallocate(edges, CAP, RES, n, res)$
  \State $degrees \gets reallocate(degrees, CAP, RES, n, res)$
  \State $capacities \gets reallocate(capacities, CAP, RES, n, res)$
  \State $\rhd$ Update vertex capacity and reserved size
  \State $CAP \gets n$ \textbf{;} $RES \gets res$
\EndFunction \label{alg:digraph--reserve-end}

\Statex

\State $\rhd$ Reallocate memory of specified size and reserved amount
\Function{reallocate}{$ptr, N_0, R_0, N_1, R_1$} \label{alg:digraph--reallocate-begin}
  \If{$R_1 = R_0$} \Comment{Desired reserved amount = old?}
    \ForAll{$i \in [N_0, N_1)$ \textbf{in parallel}} $ptr[i] \gets 0$
    \EndFor
    \Return{$ptr$}
  \EndIf
  \State $tmp \gets$ Allocate memory for $R_1$ entries
  \State $M \gets min(N_0, N_1)$
  \ForAll{$i \in [0, M)$ \textbf{in parallel}} $tmp[i] \gets ptr[i]$
  \EndFor
  \ForAll{$i \in [M, N_1)$ \textbf{in parallel}} $tmp[i] \gets 0$
  \EndFor
  \State Free memory at $ptr$
  \Return{$tmp$}
\EndFunction \label{alg:digraph--reallocate-end}
\algstore{alg:digraph12}
\end{algorithmic}
\end{algorithm}

\begin{algorithm}[hbtp]
\caption{Our Directed Graph using CP2AA allocator (Part 2).}
\label{alg:digraph2}
\begin{algorithmic}[1]
\algrestore{alg:digraph12}
\State $\rhd$ Add a vertex to the graph
\Function{addVertex}{$u$} \textbf{of} DiGraph \label{alg:digraph--add-vertex-begin}
  \If{$u \geq CAP$} $reserve(u + 1)$
  \EndIf
  \State $setBit(exists, u)$
\EndFunction \label{alg:digraph--add-vertex-end}

\Statex

\State $\rhd$ Allocate space for outgoing edges of a vertex
\Function{allocateEdges}{$u, deg$} \textbf{of} DiGraph \label{alg:digraph--allocate-edges-begin}
  \If{$u \geq CAP$ \textbf{or} $edges[u] \neq \phi$} \ReturnInline{}
  \EndIf
  \State $bytes \gets cp2aa.allocationSize(deg * \textsc{edge\_size})$
  \State $edges[u] \gets cp2aa.allocate(bytes)$
  \State $capacities[u] \gets bytes / \textsc{edge\_size}$
\EndFunction \label{alg:digraph--allocate-edges-end}

\Statex

\State $\rhd$ Add an outgoing edge, without checks
\Function{addEdgeUnsafe}{$u, v, w$} \textbf{of} DiGraph \label{alg:digraph--add-edge-unsafe-begin}
  \State $ptr \gets edges[u]$
  \State $i \gets atomicAdd(degrees[u], 1)$
  \State $ptr[i] \gets (v, w)$
\EndFunction \label{alg:digraph--add-edge-unsafe-end}

\Statex

\State $\rhd$ Add outgoing edges to a vertex
\Function{addEdges}{$u, list$} \textbf{of} DiGraph \label{alg:digraph--add-edges-begin}
  \If{\textbf{not} $hasVertex(u)$ \textbf{or} $list = \{\}$} \ReturnInline{$0$}
  \EndIf
  \State $deg_{prev} \gets degrees[u]$
  \State $deg_{max} \gets degrees[u] + size(list)$
  \State $bytes \gets cp2aa.allocationCapacity(deg_{max})$
  \State $ptr \gets cp2aa.allocate(bytes)$
  \State $degrees[u] \gets \textit{setUnion}(edges(u), list, \textbf{into}\ ptr)$
  \State $cp2aa.deallocate(edges[u], capacities[u] * \textsc{edge\_size})$
  \State $edges[u] \gets ptr$
  \State $capacities[u] \gets bytes / \textsc{edge\_size}$
  \Return{$degrees[u] - deg_{prev}$}
\EndFunction \label{alg:digraph--add-edges-end}

\Statex

\State $\rhd$ Remove outgoing edges from a vertex
\Function{removeEdges}{$u, list$} \textbf{of} DiGraph \label{alg:digraph--remove-edges-begin}
  \If{\textbf{not} $hasVertex(u)$ \textbf{or} $list = \{\}$} \ReturnInline{$0$}
  \EndIf
  \State $deg_{prev} \gets degrees[u]$
  \State $degrees[u] \gets \textit{setDifference}(\textbf{into}\ edges(u), list)$
  \Return{$deg_{prev} - degrees[u]$}
\EndFunction \label{alg:digraph--remove-edges-end}

\Statex

\State $\rhd$ Update the graph after changes
\Function{update}{$isUnique, isSorted$} \textbf{of} DiGraph \label{alg:digraph--update-begin}
  \State $\rhd$ Ensure edges are sorted and unique
  \If{\textbf{not} $isSorted$}
    \ForAll{$u \in [0, CAP)$ \textbf{in parallel}}
      \State $sortByKey(edges[u], degrees[u])$
    \EndFor
  \EndIf
  \If{\textbf{not} $isUnique$}
    \ForAll{$u \in [0, CAP)$ \textbf{in parallel}}
      \State $uniqueByKey(edges[u], degrees[u])$
    \EndFor
  \EndIf
  \State $\rhd$ Update counts of vertices and edges
  \State $N \gets M \gets 0$
  \ForAll{$u \in [0, CAP)$ \textbf{in parallel}}
    \If{\textbf{not} $getBit(exists, u)$} \textbf{continue}
    \EndIf
    \State $N \gets N + 1$ \textbf{;} $M \gets M + degrees[u]$
  \EndFor
\EndFunction \label{alg:digraph--update-end}
\end{algorithmic}
\end{algorithm}

\subsubsection{Loading a Graph from Disk}
\label{sec:load}

We now describe our algorithm for loading a graph from a Matrix Market (MTX) format file as a Compressed Sparse Row (CSR) representation. The algorithm is presented in Algorithm \ref{alg:load}, and is an improvement upon GVEL\ignore{, our previous work} \cite{sahu2023gvel}. In the algorithm, the function \texttt{loadGraph()} takes as input an MTX-formatted string, which is essentially the memory-mapped data of and MTX file, and outputs the graph $G$ in CSR format.

In the algorithm, we first initialize an empty graph structure $G$ (line \ref{alg:load--init}). We then read the MTX file header using the \texttt{readHeader()} function (line \ref{alg:load--read-header}), which indicates graph properties, such as whether the graph is symmetric, the number of rows and columns $N$, and the total number of edges $M$. If the graph is symmetric, the number of edges $M$ is doubled to account for bidirectional connections. The output CSR graph $G$ is resized accordingly to accommodate the required storage space (line \ref{alg:load--resize}).
Next, we allocate space for storing edge data in parallel (lines \ref{alg:load--alloc-edges-begin}-\ref{alg:load--alloc-edges-end}). Each thread is responsible for allocating memory for $source$ and $target$ vertices, and $weights$, if the graph is weighted. The collection of allocated edge lists is stored in the variable $edges$ (line \ref{alg:load--edges}). Next, space is allocated for per-partition degree counts $pdegrees$ and per-partition CSRs $pcsr$, which includes per-partition offsets, edge keys, and edge values (lines \ref{alg:load--alloc-pdegrees-pcsr-begin}-\ref{alg:load--alloc-pdegrees-pcsr-end}). The first partition ($p=0$) is mapped directly to the primary degree and CSR arrays of $G$. For subsequent partitions ($p \in [1, \rho)$), where $\rho$ is the number of partitions, space is allocated separately. If the graph is weighted, additional memory is assigned for storing edge weights. Each partition, the degree array is initialized to zero to prepare for counting.

After setting up the necessary data structures, we read the edge list and processes it into CSR format (line \ref{alg:load--read-edgelist}-\ref{alg:load--convert-to-csr}). The function \texttt{readEdgelist()} populates the per-thread edges and per-partition degree counts, while \texttt{convertToCsr()} converts this data into a CSR representation. For this, \texttt{convertToCsr()} utilizes $\rho$ per-partition CSRs ($pcsr$) as intermediates, in order to generate the global CSR $G$. If the graph is symmetric, the $G$ is resized (line \ref{alg:load--resize-symmetric}). Finally, we return the constructed CSR graph $G$ (line \ref{alg:load--return}).

\begin{algorithm}[hbtp]
\caption{Load a graph from an MTX format file as a CSR.}
\label{alg:load}
\begin{algorithmic}[1]
\Require{$data$: Input MTX data as string}
\Ensure{$G$: Output CSR graph}
\Ensure{$symmetric$: Is graph symmetric?}
\Ensure{$weighted$: Is graph weighted?}
\Ensure{$\rho$: Number of partitions for counting vertex degrees}
\Ensure{$t$: Current thread}

\Statex

\Function{loadGraph}{$data$}
  \State $G \gets \{\}$ \label{alg:load--init}
  \State $\rhd$ Read MTX format header
  \State $(symmetric, rows, cols, size, head) \gets readHeader(data)$ \label{alg:load--read-header}
  \State $data \gets removePrefix(data, head)$
  \State $\rhd$ Allocate space for CSR
  \State $N \gets max(rows, cols)$
  \State $M \gets 2 * size$ \textbf{if} $symmetric$ \textbf{else} $size$
  \State $G.resize(N, M)$ \label{alg:load--resize}
  \State $\rhd$ Allocate space for edges
  \State $sources \gets targets \gets weights \gets \{\}$ \label{alg:load--alloc-edges-begin}
  \ForAll{\textbf{threads in parallel}}
      \State $sources[t] \gets$ Allocate space for $M$ elements
      \State $targets[t] \gets$ Allocate space for $M$ elements
      \If{$weighted$}
          \State $weights[t] \gets$ Allocate space for $M$ elements
      \EndIf
  \EndFor \label{alg:load--alloc-edges-end}
  \State $edges \gets (sources, targets, weights)$ \label{alg:load--edges}
  \State $\rhd$ Allocate space for pdegrees, pcsr
  \State $pdegrees \gets \textit{poffsets} \gets pedgeKeys \gets \textit{pedgeValues} \gets \{\}$ \label{alg:load--alloc-pdegrees-pcsr-begin}
  \State $pdegrees[0] \gets G.degrees$
  \State $\textit{poffsets}[0] \gets G.offsets$
  \State $pedgeKeys[0] \gets G.edgeKeys$
  \If{$weighted$}
      \State $\textit{pedgeValues}[0] \gets G.edgeValues$
  \EndIf
  \State $fill(pdegrees[0], 0)$
  \ForAll{$p \in [1, \rho)$ \textbf{in parallel}}
      \State $pdegrees[p] \gets$ Allocate space for $N$ elements
      \State $\textit{poffsets}[p] \gets$ Allocate space for $N+1$ elements
      \State $pedgeKeys[p] \gets$ Allocate space for $M$ elements
      \If{$weighted$}
          \State $\textit{pedgeValues}[p] \gets$ Allocate space for $M$ elements
      \EndIf
      \State $fill(pdegrees[p], 0)$
  \EndFor \label{alg:load--alloc-pdegrees-pcsr-end}
  \State $pcsr \gets (\textit{poffsets}, pedgeKeys, \textit{pedgeValues})$
  \State $\rhd$ Read edge list and convert to CSR
  \State $counts \gets readEdgelist(pdegrees, edges, data)$ \label{alg:load--read-edgelist}
  \State $M \gets convertToCsr(pcsr, pdegrees, edges, counts)$ \label{alg:load--convert-to-csr}
  \State $\rhd$ Account for self-loops in symmetric graphs
  \If{$symmetric$} $G.resize(N, M)$ \label{alg:load--resize-symmetric}
  \EndIf
  \Return{$G$} \label{alg:load--return}
\EndFunction
\end{algorithmic}
\end{algorithm}

We now discuss the psuedocode of the \texttt{readEdgelist()} function, which is described in Algorithm \ref{alg:load-el}. \texttt{readEdgelist()} takes as input the per-partition vertex degrees, $pdegrees$; an output structure for storing edges, $edges$; and the memory-mapped file data, $data$. The function processes the edgelist in parallel, in blocks of size $\beta$. The parsed edges are populated, and the number of edges read per thread is tracked using $counts$, which is returned.

In the function, we begin by initializing the count of edges processed per thread and extracting the per-thread sources, targets, and weights of edges (lines \ref{alg:el--initialize-begin}-\ref{alg:el--initialize-end}). Next, edges are loaded from the file in blocks of size $\beta$, ensuring efficient reading of large files (lines \ref{alg:el--blocks-begin}-\ref{alg:el--blocks-end}). Each thread processes a specific range of the input file, determined by its index in steps of $\beta$. The function \texttt{getBlock()} extracts the corresponding block from the memory-mapped file (line \ref{alg:el--get-block}). Within each block, edges are read sequentially (lines \ref{alg:el--block-begin}-\ref{alg:el--block-end}). Each edge consists of a pair of vertex identifiers $(u, v)$, with an optional weight $w$ if the graph is weighted (lines \ref{alg:el--parse-edge-begin}-\ref{alg:el--parse-edge-end}). The edge data is parsed by first locating the next numeric digit in the file, extracting the integer values for $u$ and $v$, and, if applicable, parsing a floating-point weight. The vertex identifiers are then converted to zero-based indexing (line \ref{alg:el--base1}). Once an edge is parsed, it is added to the corresponding thread's edge list (lines \ref{alg:el--add-edge-begin}-\ref{alg:el--add-edge-end}). Additionally, the per-partition vertex degree counter is updated atomically to reflect the addition of the edge (line \ref{alg:el--update-degrees}). If the graph is symmetric, the reverse edge $(v, u)$ is also stored, ensuring undirected connectivity is maintained (lines \ref{alg:el--reverse-edge-begin}-\ref{alg:el--reverse-edge-end}). After processing all edges in a block, the count of edges processed by each thread is updated (line \ref{alg:el--update-counts}). Once all blocks are processed, we return the total number of edges read per thread\ignore{(line \ref{alg:el--return-counts})}.

The \texttt{getBlock()} function ensures that each thread correctly extracts a contiguous chunk of the file while respecting line boundaries (lines \ref{alg:el--get-block-begin}-\ref{alg:el--get-block-end}). It first determines the starting position $b$ and the endpoint $B$ of the block within the file. If the starting position is within a partially read line, it adjusts $b$ to the beginning of the next line. Similarly, if the endpoint $B$ is within a partially read line, it extends $B$ to include the full line. The function then returns the adjusted block boundaries $[b, B]$, ensuring that each thread reads a well-formed subset of the edgelist.

\begin{algorithm}[hbtp]
\caption{Reading Edgelist from file.}
\label{alg:load-el}
\begin{algorithmic}[1]
\Require{$pdegrees$: Per partition vertex degrees (output)}
\Require{$edges$: Per thread sources, targets, and weights of edges (output)}
\Require{$data$: Memory mapped file data}
\Ensure{$counts$: Number of edges read per thread (output)}
\Ensure{$symmetric$: Is graph symmetric?}
\Ensure{$weighted$: Is graph weighted?}
\Ensure{$\beta$: Size of each block that is processed per thread}
\Ensure{$\rho$: Number of partitions for counting vertex degrees}
\Ensure{$t$: Current thread}

\Statex

\Function{readEdgelist}{$pdegrees, edges, data$} \label{alg:el--read-edgelist-begin}
  \State $counts \gets \{0\}$ \label{alg:el--initialize-begin}
  \State $(sources, targets, weights) \gets edges$ \label{alg:el--initialize-end}
  \State $\rhd$ Load edges from text file in blocks of size $\beta$
  \ForAll{$i \in [0, \beta, 2\beta, ... |data|]$ \textbf{in parallel}} \label{alg:el--blocks-begin}
    \State $j \gets counts[t]$
    \State $[b, B] \gets getBlock(data, i)$ \label{alg:el--get-block}
    \While{$true$} \label{alg:el--block-begin}
      \State $\rhd$ Read an edge from the block
      \State $u \gets v \gets 0$ \textbf{;} $w \gets 1$ \label{alg:el--parse-edge-begin}
      \State $b \gets findNextDigit(b, B)$
      \If{$b = B$} \textbf{break}
      \EndIf
      \State $b \gets parseWholeNumber(u, b, B)$
      \State $b \gets findNextDigit(b, B)$
      \State $b \gets parseWholeNumber(v, b, B)$
      \If{$weighted$}
        \State $b \gets findNextDigit(b, B)$
        \State $b \gets parseFloat(w, b, B)$
      \EndIf \label{alg:el--parse-edge-end}
      \State $\rhd$ Make it zero-based
      \State $u \gets u - 1$ \textbf{;} $v \gets v - 1$ \label{alg:el--base1}
      \State $\rhd$ Add the parsed edge to edgelist
      \State $sources[t][j] \gets u$ \label{alg:el--add-edge-begin}
      \State $targets[t][j] \gets v$
      \If{$weighted$} $weights[t][j] \gets w$
      \EndIf
      \State $atomicAdd(pdegrees[t \bmod \rho][u], 1)$ \label{alg:el--update-degrees}
      \State $j \gets j + 1$ \label{alg:el--add-edge-end}
      \State $\rhd$ If graph is symmetric, add the reverse edge
      \If{$symmetric$} \label{alg:el--reverse-edge-begin}
        \State $sources[t][j] \gets v$
        \State $targets[t][j] \gets u$
        \If{$weighted$} $weights[t][j] \gets w$
        \EndIf
        \State $atomicAdd(pdegrees[t \bmod \rho][v], 1)$
        \State $j \gets j + 1$
      \EndIf \label{alg:el--reverse-edge-end}
    \EndWhile \label{alg:el--block-end}
    \State $counts[t] \gets j$ \label{alg:el--update-counts}
  \EndFor \label{alg:el--blocks-end}
  \Return{$counts$} \label{alg:el--return-counts}
\EndFunction \label{alg:el--read-edgelist-end}

\Statex

\Function{getBlock}{$data, i$} \label{alg:el--get-block-begin}
  \State $[d, D] \gets data$
  \State $b \gets d+i$ \textbf{;} $B \gets min(b+\beta, D)$
  \If{$b \neq d$ \textbf{and not} $isNewline(b-1)$}
    \State $b \gets findNextLine(b, D)$
  \EndIf
  \If{$B \neq d$ \textbf{and not} $isNewline(B-1)$}
    \State $B \gets findNextLine(B, D)$
  \EndIf
  \Return{$[b, B]$}
\EndFunction \label{alg:el--get-block-end}
\end{algorithmic}
\end{algorithm}

Next, we discuss the pseudocode of the \texttt{convertToCsr()} function, which is given in Algorithm \ref{alg:load-csr}. \texttt{convertToCsr()} takes as input the per-partition compressed sparse row (CSR) structure, $pcsr$; per-partition vertex degrees, $pdegrees$; the per-thread edge lists containing sources, targets, and weights of edges, $edges$; and the number of edges read per thread, $counts$. Our goal is to efficiently convert the per-thread edgelists into a partitioned CSR representation and then merge these partitioned CSRs into a final global CSR. By partitioning the CSR construction, we minimize contention when updating global data structures, significantly improving performance compared to a direct global CSR construction.

We begin by extracting the key components of the per-partition CSR: offsets ($poffsets$), edge destinations ($pedgeKeys$), and edge weights ($pedgeValues$) (lines \ref{alg:csr--initialize-begin}-\ref{alg:csr--initialize-end}). Additionally, we retrieve the per-thread sources, targets, and weights of edges. Next, we compute the global degree of each vertex and store it in $degrees[0]$. This step is essential because we use partition $0$ as the base for the final CSR, ensuring that its offsets are correctly initialized. Although we construct the CSR in four partitions ($0$ to $\rho-1$), partition $0$ is already properly set up, allowing us to merge only the remaining three partitions into it (lines \ref{alg:csr--initialize-end}-\ref{alg:csr--poffsets-begin}).
To construct the per-partition CSR efficiently, we compute shifted offsets for each partition (lines \ref{alg:csr--poffsets-begin}-\ref{alg:csr--poffsets-end}). This optimization ensures that offsets can be used directly as indices for inserting edges into the CSR. Once the CSR is populated, the shifted offsets will automatically reflect the final correct offsets without requiring additional post-processing steps, which used to be necessary in older approaches such as GVEL \cite{sahu2023gvel}. This avoids an extra pass over the data to adjust the offsets.

We then populate the per-partition CSR in parallel (lines \ref{alg:csr--pcsr-begin}-\ref{alg:csr--pcsr-end}). Each thread processes its assigned edges and determines which partition they belong to. For each edge $(u, v)$, we atomically increment the corresponding offset and store $v$ in $pedgeKeys$. If the graph is weighted, we also store the corresponding weight in $pedgeValues$. Partitioning the CSR generation in this manner significantly reduces contention on atomic operations, as each partition is updated independently by different sets of threads.
After constructing the per-partition CSRs, we merge them into a single CSR (lines \ref{alg:csr--pcsr-combine-begin}-\ref{alg:csr--pcsr-combine-end}). Since partition $0$ is already structured correctly, we only need to merge partitions $1$ to $\rho-1$ into it. This process iterates over all vertices and sequentially appends their edges from partitions $1$ to $\rho-1$ into partition $0$. As edges are copied, we maintain an index $j$ to track the insertion position in the merged CSR. If the graph is weighted, we also copy the corresponding weights. Finally, we update the offsets in partition $0$ to reflect the total number of edges per vertex, completing the global CSR construction, and return the total number of edges in the CSR, $M$.

This algorithm differs from the one used in GVEL \cite{sahu2023gvel} in several key ways. First, we compute the global degree of each vertex in $degrees[0]$ to prepare partition $0$ as the final CSR, reducing the complexity of the merging step. Second, we use shifted offsets during CSR initialization, eliminating the need for a post-processing step to fix offsets after populating the CSR.

\begin{algorithm}[hbtp]
\caption{Convert per-thread Edgelists to CSR.}
\label{alg:load-csr}
\begin{algorithmic}[1]
\Require{$pcsr$: Per partition CSR (scratch)}
\Require{$pdegrees$: Per partition vertex degrees (scratch)}
\Require{$edges$: Per thread sources, targets, and weights of edges}
\Require{$counts$: Number of edges read per thread}
\Ensure{$symmetric$: Is graph symmetric?}
\Ensure{$weighted$: Is graph weighted?}
\Ensure{$\rho$: Number of partitions for counting vertex degrees}
\Ensure{$t$: Current thread}

\Statex

\Function{convertToCsr}{$pcsr, pdegrees, edges, counts$}
  \State $(\textit{poffsets}, pedgeKeys, pedgeValues) \gets pcsr$ \label{alg:csr--initialize-begin}
  \State $(sources, targets, weights) \gets edges$ \label{alg:csr--initialize-end}
  \State $\rhd$ Compute global degrees at $degrees[0]$
  \ForAll{$u \in [0, |V|)$ \textbf{in parallel}}
    \ForAll{$p \in [1, \rho)$} $degrees[0]\ \text{+=}\ degrees[p]$
    \EndFor
  \EndFor
  \State $\rhd$ Compute per-partition shifted offsets
  \ForAll{$p \in [0, \rho)$} \label{alg:csr--poffsets-begin}
    \State $M_p \gets exclusiveScan(\textit{poffsets}[p]+1, \textit{pdegrees}[p], |V|)$
    \If{$p = 0$} $M \gets M_p$
    \EndIf
  \EndFor \label{alg:csr--poffsets-end}
  \State $\rhd$ Populate per-partition CSR
  \ForAll{\textbf{threads in parallel}} \label{alg:csr--pcsr-begin}
    \State $p \gets t \bmod \rho$
    \ForAll{$i \in [0, counts[t])$}
      \State $u \gets sources[t][i]$
      \State $v \gets targets[t][i]$
      \State $j \gets atomicAdd(\textit{poffsets}[p][u+1], 1)$
      \State $pedgeKeys[p][j] \gets v$
      \If{$weighted$}
        \State $pedgeValues[p][j] \gets weights[t][i]$
      \EndIf
    \EndFor
  \EndFor \label{alg:csr--pcsr-end}
  \State $\rhd$ Combine per-partition CSR into one CSR
  \ForAll{$u \in [0, |V|)$ \textbf{in parallel}} \label{alg:csr--pcsr-combine-begin}
    \State $j \gets \textit{poffsets}[0][u+1]$
    \ForAll{$p \in [1, \rho)$}
      \State $i \gets \textit{poffsets}[p][u]$
      \State $I \gets \textit{poffsets}[p][u+1]$
      \ForAll{$i \in [i, I)$}
        \State $pedgeKeys[0][j] \gets pedgeKeys[p][i]$
        \If{$weighted$}
          \State $pedgeValues[0][j] \gets pedgeValues[p][i]$
        \EndIf
        \State $j \gets j + 1$
      \EndFor
    \EndFor
    \State $\textit{poffsets}[0][u+1] \gets j$
  \EndFor \label{alg:csr--pcsr-combine-end}
  \Return{$M$}
\EndFunction
\end{algorithmic}
\end{algorithm}

\subsubsection{Cloning a Graph}
\label{sec:clone}

The \texttt{cloneGraph()} function we use for (parallel) cloning of our graph representation is shown in Algorithm \ref{alg:clone}. It takes an input graph $G$, and returns the cloned graph $G'$.

The method begins by initializing an empty graph $G'$ (line \ref{alg:clone--init}). To optimize memory allocation, we preallocate storage for vertices based on the maximum vertex ID in $G$ (line \ref{alg:clone--reserve-vertices}). Next, all vertices from $G$ are added to $G'$ in parallel (lines \ref{alg:clone--add-vertices-begin}-\ref{alg:clone--add-vertices-end}), ensuring that all nodes from $G$ are present in $G'$ before edges are processed. Following this, we reserve space for edges by preallocating memory for each vertex’s adjacency list (lines \ref{alg:clone--reserve-edges-begin}-\ref{alg:clone--reserve-edges-end}). Once vertices are inserted, we preallocate edge storage for each vertex $u$ in parallel, setting aside space based on its degree in $G$ (lines \ref{alg:clone--populate-edges-begin}-\ref{alg:clone--populate-edges-end}). Next, the edges linked to vertex $u \in V$, and its associated degree are directly copied from the original graph. Finally, we update the total vertex and edge counts in $G'$. If $G$ is stored in Compressed Sparse Row (CSR) format, a specialized \texttt{update()} function is called to additionally sort the edges of each vertex in the graph by ID (line \ref{alg:clone--update-counts-csr}) --- this is done to ensure that edge additions and insertions can be done in $O(d_u + \Delta d_u)$ time, where $d_u$ is the degree of vertex $u$, and $\Delta d_u$ is the number of edges removed are being added to the vertex. Otherwise, the vertex and edge counts are explicitly set (line \ref{alg:clone--update-counts-noncsr}). The cloned graph $G'$, which is a deep-copy of $G$, is returned in line \ref{alg:clone--return}.

\begin{algorithm}[hbtp]
\caption{Create a deep-copy of a graph.}
\label{alg:clone}
\begin{algorithmic}[1]
\Require{$G(V, E)$: Input graph}
\Require{$G'(V', E')$: Cloned graph}

\Statex

\Function{cloneGraph}{$G$}
  \State $G' \gets \{\}$ \label{alg:clone--init}
  \State $\rhd$ Reserve space for vertices
  \State $G'.reserve(maxVertexId(G) + 1)$ \label{alg:clone--reserve-vertices}
  \State $\rhd$ Add vertices
  \ForAll{$u \in V$ \textbf{in parallel}} \label{alg:clone--add-vertices-begin}
    \State $G'.addVertex(u)$
  \EndFor \label{alg:clone--add-vertices-end}
  \State $\rhd$ Reserve space for edges
  \ForAll{$u \in V$ \textbf{in parallel}} \label{alg:clone--reserve-edges-begin}
    \State $G'.allocateEdges(u, G.degree(u))$
  \EndFor \label{alg:clone--reserve-edges-end}
  \State $\rhd$ Populate the edges
  \ForAll{$u \in V$ \textbf{in parallel}} \label{alg:clone--populate-edges-begin}
    \State $G'.edges(u) \gets G.edges(u)$
    \State $G'.degree(u) \gets G.degree(u)$
  \EndFor \label{alg:clone--populate-edges-end}
  \State $\rhd$ Update vertex and edge counts
  \If{$G.isCsr()$} $G'.update(true, false)$ \label{alg:clone--update-counts-csr}
  \Else\ $|V'| \gets |V|$ \textbf{;} $|E'| \gets |E|$ \label{alg:clone--update-counts-noncsr}
  \EndIf
  \Return{$G'$} \label{alg:clone--return}
\EndFunction
\end{algorithmic}
\end{algorithm}

\subsubsection{Performing Edge Deletions}
\label{sec:sub}

We now discuss our graph subtraction algorithms, which remove the edges of a graph $G_S$ from another graph $G$, effectively applying a batch edge deletions $E_S$ to $G$. The pseudocode is provided in Algorithm \ref{alg:sub}. Algorithm \ref{alg:sub} includes two functions: \textbf{(1)} \texttt{subtractGraphInplace()} modifies $G$ directly, removing edges in $G_S$ from $G$. \textbf{(2)} \texttt{subtractGraph()} returns a new graph $G'$ such that $G' = G \setminus G_S$ --- it is more efficient than simply cloning $G$, and then performing the graph subtraction in-place.

In \texttt{subtractGraphInplace()}, we initialize the count of removed edges, $\Delta M$, to zero. Next, we iterate over all vertices in $G_S$ in parallel (line \ref{alg:sub--inplace-for-begin}). If a vertex $u$ from $G_S$ is not present in $G$, it is skipped (line \ref{alg:sub--inplace-check}). Otherwise, we remove the edges of $u$ that are present in $G_S$ from $G$, updating $\Delta M$ accordingly (line \ref{alg:sub--inplace-remove}). Once all relevant edges have been deleted, the total edge count of $G$ is updated (line \ref{alg:sub--inplace-update-edges}), and the modified graph is returned (line \ref{alg:sub--inplace-return}). This function operates in-place, meaning the original graph $G$ is directly modified.

The \texttt{subtractGraph()} function constructs a new graph $G'$ instead of modifying $G$ directly. First, an empty graph $G'$ is initialized, and $\Delta M$ is set to zero (line \ref{alg:sub--init}). To ensure efficient memory allocation, we reserve space for the vertices in $G'$ (line \ref{alg:sub--reserve-vertices}). The vertices of $G$ are then added to $G'$ in parallel (lines \ref{alg:sub--add-vertices-begin}-\ref{alg:sub--add-vertices-end}). Similarly, memory is allocated for the edges of each vertex, based on its degree in $G$ (lines \ref{alg:sub--alloc-edges-begin}-\ref{alg:sub--alloc-edges-end}). The edge copying process is performed in two steps. First, for vertices that are not present in $G_S$, all edges from $G$ are copied directly into $G'$ (lines \ref{alg:sub--copy-untouched-begin}-\ref{alg:sub--copy-untouched-end}). This ensures that the structure of these vertices remains unchanged. Next, for vertices that do exist in $G_S$, we remove the edges that appear in $G_S$, keeping only those that are not part of $G_S$ (lines \ref{alg:sub--copy-touched-begin}-\ref{alg:sub--copy-touched-end}). The degree of each vertex is updated accordingly, and $\Delta M$ is incremented based on the number of removed edges. Finally, the vertex and edge counts for $G'$ are updated (line \ref{alg:sub--update-counts}), and the new graph $G'$ is returned (line \ref{alg:sub--return}).

\begin{algorithm}[hbtp]
\caption{Subtract a graph's edges from another graph.}
\label{alg:sub}
\begin{algorithmic}[1]
\Require{$G(V, E)$: Input graph to subtract from}
\Require{$G_S(V_S, E_S)$: Graph containing the edges to subtract}
\Require{$G'(V', E')$: Output graph, with edges from $G_S$ removed}
\Ensure{$\Delta M$: Number of edges deleted from $G$}

\Statex

\Function{subtractGraphInplace}{$G, G_S$}
  \State $\Delta M \gets 0$
  \State $\rhd$ Remove the edges in-place
  \ForAll{$u \in V_S$ \textbf{in parallel}} \label{alg:sub--inplace-for-begin}
    \If{\textbf{not} $G.hasVertex(u)$} \textbf{continue} \label{alg:sub--inplace-check}
    \EndIf
    \State $\Delta M \gets \Delta M + G.removeEdges(u, G_S.edges(u))$ \label{alg:sub--inplace-remove}
  \EndFor \label{alg:sub--inplace-for-end}
  \State $\rhd$ Update the edge count
  \State $|E| \gets |E| - \Delta M$ \label{alg:sub--inplace-update-edges}
  \Return{$G$} \label{alg:sub--inplace-return}
\EndFunction

\Statex

\Function{subtractGraph}{$G, G_S$}
  \State $G' \gets \{\}$ \textbf{;} $\Delta M \gets 0$ \label{alg:sub--init}
  \State $\rhd$ Reserve space for vertices
  \State $G'.reserve(maxVertexId(G) + 1)$ \label{alg:sub--reserve-vertices}
  \State $\rhd$ Add vertices
  \ForAll{$u \in V$ \textbf{in parallel}} \label{alg:sub--add-vertices-begin}
    \State $G'.addVertex(u)$
  \EndFor \label{alg:sub--add-vertices-end}
  \State $\rhd$ Reserve space for edges
  \ForAll{$u \in V$ \textbf{in parallel}} \label{alg:sub--alloc-edges-begin}
    \State $G'.allocateEdges(u, G.degree(u))$
  \EndFor \label{alg:sub--alloc-edges-end}
  \State $\rhd$ Add edges of vertices that are untouched
  \ForAll{$u \in V$ \textbf{in parallel}} \label{alg:sub--copy-untouched-begin}
    \If{$G_S.hasVertex(u)$} \textbf{continue}
    \EndIf
    \State $G'.edges(u) \gets G.edges(u)$
    \State $G'.degree(u) \gets G.degree(u)$
  \EndFor \label{alg:sub--copy-untouched-end}
  \State $\rhd$ Add edges of vertices that are touched
  \ForAll{$u \in V$ \textbf{in parallel}} \label{alg:sub--copy-touched-begin}
    \If{\textbf{not} $G_S.hasVertex(u)$} \textbf{continue}
    \EndIf
    \State $G'.edges(u) \gets G.edges(u) \setminus G_S.edges(u)$
    \State $G'.degree(u) \gets |G'.edges(u)|$
    \State $\Delta M \gets \Delta M + |G.edges(u)| - |G'.edges(u)|$
  \EndFor \label{alg:sub--copy-touched-end}
  \State $\rhd$ Update the vertex and edge count
  \State $|V'| \gets |V|$ \textbf{;} $|E'| \gets |E| - \Delta M$ \label{alg:sub--update-counts}
  \Return{$G'$} \label{alg:sub--return}
\EndFunction
\end{algorithmic}
\end{algorithm}

\subsubsection{Performing Edge Insertions}
\label{sec:add}

Next, we describe our graph union algorithms, detailed in Algorithm \ref{alg:add}, which merges the edges of a graph $G_A$ into another graph $G$, effectively applying a batch of edge insertions $E_A$ to $G$. This done either by modifying $G$ in place (in the \texttt{addGraphInplace()} function), or by generating a new output graph $G'$ (in the \texttt{addGraph()} function).

In the first function, \texttt{addGraphInplace()}, we modify $G$ directly. Initially, the counters $\Delta N$ and $\Delta M$ are set to zero, representing the number of new vertices and edges added, respectively (line \ref{alg:add--inplace-init}). We then allocate space for vertices by reserving memory for $max(\text{vertex ID in } G, \text{vertex ID in } G_A) + 1$\ignore{(line \ref{alg:add--inplace-reserve})}. Next, new vertices from $G_A$ that are not already in $G$ are added in parallel, updating $\Delta N$ accordingly (lines \ref{alg:add--add-vertices-begin}-\ref{alg:add--add-vertices-end}). After this, edges from $G_A$ are inserted into $G$ in-place (lines \ref{alg:add--inplace-add-edges-begin}-\ref{alg:add--inplace-add-edges-end}), using the \texttt{G.addEdges()} function detailed in Algorithm \ref{alg:digraph2}, and the count of newly added edges is accumulated in $\Delta M$. Finally, the vertex and edge counts of $G$ are updated to reflect the additions (line \ref{alg:add--inplace-update}), and the modified graph $G$ is returned\ignore{(line \ref{alg:add--inplace-return})}.

The second function, \texttt{addGraph()}, constructs a new graph $G'$ containing the union of graphs $G$ and $G_A$. The function starts by initializing $G'$ as an empty graph and setting $\Delta N$ and $\Delta M$ to zero (line \ref{alg:add--init}). Similar to the in-place version, we allocate space for vertices (line \ref{alg:add--reserve}) and then iterate through the vertex range to add vertices from $G$ and $G_A$, tracking the count of new additions in $\Delta N$ (lines \ref{alg:add--add-vertices-begin}-\ref{alg:add--add-vertices-end}). After reserving memory for edges (lines \ref{alg:add--reserve-edges-begin}-\ref{alg:add--reserve-edges-end}), we proceed to populate $G'$ with edges. First, we copy edges for vertices that exist only in $G$ (lines \ref{alg:add--add-edges-untouched-begin}-\ref{alg:add--add-edges-untouched-end}). Then, for vertices present in both $G$ and $G_A$, we merge their edge sets and update the degree count (lines \ref{alg:add--add-edges-touched-begin}-\ref{alg:add--add-edges-touched-end}). The total number of newly added edges is accumulated in $\Delta M$. Finally, the vertex and edge counts for $G'$ are updated (line \ref{alg:add--update}), and the new graph is returned (line \ref{alg:add--return}).

\begin{algorithm}[hbtp]
\caption{Add a graph's edges to another graph.}
\label{alg:add}
\begin{algorithmic}[1]
\Require{$G(V, E)$: Input graph to add to}
\Require{$G_A(V_A, E_A)$: Graph containing the edges to add}
\Require{$G'(V', E')$: Output graph, with edges from $G_A$ added}
\Ensure{$\Delta N$: Number of vertices added from $G_A$}
\Ensure{$\Delta M$: Number of edges inserted from $G_A$}

\Statex

\Function{addGraphInplace}{$G, G_A$}
  \State $\Delta N \gets \Delta M \gets 0$ \label{alg:add--inplace-init}
  \State $\rhd$ Reserve space for vertices
  \State $G.reserve(maxVertexId(G, G_A) + 1)$ \label{alg:add--inplace-reserve}
  \State $\rhd$ Add new vertices
  \ForAll{$u \in G_A$ \textbf{in parallel}} \label{alg:add--inplace-add-vertices-begin}
    \If{$G.hasVertex(u)$} \textbf{continue}
    \EndIf
    \State $G.addVertex(u)$ \textbf{;} $\Delta N \gets \Delta N + 1$
  \EndFor \label{alg:add--inplace-add-vertices-end}
  \State $\rhd$ Add the edges in-place
  \ForAll{$u \in V$ \textbf{in parallel}} \label{alg:add--inplace-add-edges-begin}
    \State $\Delta M \gets \Delta M + G.addEdges(u, G_A.edges(u))$
  \EndFor \label{alg:add--inplace-add-edges-end}
  \State $\rhd$ Update the vertex and edge count
  \State $|V| \gets |V| + \Delta N$ \textbf{;} $|E| \gets |E| + \Delta M$ \label{alg:add--inplace-update}
  \Return{$G$} \label{alg:add--inplace-return}
\EndFunction

\Statex

\Function{addGraph}{$G, G_A$}
  \State $G' \gets \{\}$ \textbf{;} $\Delta N \gets \Delta M \gets 0$ \label{alg:add--init}
  \State $\rhd$ Reserve space for vertices
  \State $G'.reserve(maxVertexId(G, G_A) + 1)$ \label{alg:add--reserve}
  \State $\rhd$ Add vertices
  \ForAll{$u \in [0, maxVertexId(G, G_A)]$ \textbf{in parallel}} \label{alg:add--add-vertices-begin}
    \If{$G.hasVertex(u)$}
      \State $G'.addVertex(u)$
    \ElsIf{$G_A.hasVertex(u)$}
      \State $G'.addVertex(u)$ \textbf{;} $\Delta N \gets \Delta N + 1$
    \EndIf
  \EndFor \label{alg:add--add-vertices-end}
  \State $\rhd$ Reserve space for edges
  \ForAll{$u \in V'$ \textbf{in parallel}} \label{alg:add--reserve-edges-begin}
    \State $G'.allocateEdges(u, G.degree(u) + G_A.degree(u))$
  \EndFor \label{alg:add--reserve-edges-end}
  \State $\rhd$ Add edges of vertices that are untouched
  \ForAll{$u \in V'$ \textbf{in parallel}} \label{alg:add--add-edges-untouched-begin}
    \If{$G_A.hasVertex(u)$} \textbf{continue}
    \EndIf
    \State $G'.edges(u) \gets G.edges(u)$
    \State $G'.degree(u) \gets G.degree(u)$
  \EndFor \label{alg:add--add-edges-untouched-end}
  \State $\rhd$ Add edges of vertices that are touched
  \ForAll{$u \in V'$ \textbf{in parallel}} \label{alg:add--add-edges-touched-begin}
    \If{\textbf{not} $G.hasVertex(u)$} \textbf{continue}
    \EndIf
    \State $G'.edges(u) \gets G.edges(u) \cup G_A.edges(u)$
    \State $G'.degree(u) \gets |G'.edges(u)|$
    \State $\Delta M \gets \Delta M + |G'.edges(u)| - |G.edges(u)|$
  \EndFor \label{alg:add--add-edges-touched-end}
  \State $\rhd$ Update the vertex and edge count
  \State $|V'| \gets |V| + \Delta N$ \textbf{;} $|E'| \gets |E| + \Delta M$ \label{alg:add--update}
  \Return{$G'$} \label{alg:add--return}
\EndFunction
\end{algorithmic}
\end{algorithm}

\section{Evaluation}
\label{sec:evaluation}
\subsection{Experimental Setup}
\label{sec:setup}

\subsubsection{System used}
\label{sec:system}

We utilize a server featuring two Intel Xeon Gold 6226R processors, each with 16 cores clocked at 2.90 GHz, along with 512 GB of RAM. Each core has a 1 MB L1 cache, a 16 MB L2 cache, and a 22 MB shared L3 cache. This system runs CentOS Stream 8. For GPU-based evaluations of cuGraph, we use a separate server equipped with an NVIDIA A100 GPU, which has 108 SMs with 64 CUDA cores each, 80 GB of global memory, a bandwidth of 1935 GB/s, and 164 KB of shared memory per SM. This server has an AMD EPYC-7742 processor with 64 cores running at 2.25 GHz, 512 GB of DDR4 RAM, and Ubuntu 20.04.

\subsubsection{Configuration}
\label{sec:configuration}

We use 32-bit integers for vertex IDs. For compilation, we employ GCC 13.2 and OpenMP 5.0 on the CPU-only system, while on the GPU system, we use GCC 9.4, OpenMP 5.0, and CUDA 11.4. Our CP2AA allocator, used in the DiGraph, has a pool size of $512$ KB and handles allocations between $16$ and $8192$ bytes, with other sizes routed to \texttt{new[]}/\texttt{delete[]}. When reading graphs, each thread processes text in $256$ KB blocks and uses four partitions when converting per-thread Edgelist to CSR. Most parallel operations leverage OpenMP dynamic scheduling with chunk sizes of $512$, $1024$, or $2048$. For DiGraph updates (see \texttt{update()} in Algorithm \ref{alg:digraph2}), we employ task-based parallelism in OpenMP for vertices with a degree greater than $2048$.

\subsubsection{Dataset}
\label{sec:dataset}

The graphs used in our experiments are listed in Table \ref{tab:dataset}, and they are sourced from the SuiteSparse Matrix Collection \cite{suite19}. The graphs vary in size, with the number of vertices ranging from $3.07$ million to $214$ million, and the number of edges ranging from $25.4$ million to $3.80$ billion. We ensure that the edges are undirected and weighted, with a default weight of $1$.

\begin{table}[hbtp]
  \centering
  \caption{List of $12$ graphs obtained from the SuiteSparse Matrix Collection \cite{suite19} (with directed graphs marked by $*$). Here, $|V|$ represents the number of vertices, $|E|$ the number of edges, $D_{avg}$ the average degree, and $t_{load}$ the time to load the graph from an MTX file into CSR format using Algorithm \ref{alg:load}.}
  \label{tab:dataset}
  \begin{tabular}{|c||c|c|c|c|}
    \toprule
    \textbf{Graph} &
    \textbf{\textbf{$|V|$}} &
    \textbf{\textbf{$|E|$}} &
    \textbf{\textbf{$D_{avg}$}} &
    \textbf{\textbf{$t_{load}$}} \\
    \midrule
    \multicolumn{5}{|c|}{\textbf{Web Graphs (LAW)}} \\ \hline
    indochina-2004$^*$ & 7.41M & 194M & 26.2 & 0.50s \\ \hline
    arabic-2005$^*$ & 22.7M & 640M & 28.1 & 1.34s \\ \hline
    uk-2005$^*$ & 39.5M & 936M & 23.7 & 1.70s \\ \hline
    webbase-2001$^*$ & 118M & 1.02B & 8.6 & 1.90s \\ \hline
    it-2004$^*$ & 41.3M & 1.15B & 27.9 & 2.37s \\ \hline 
    sk-2005$^*$ & 50.6M & 1.95B & 38.5 & 5.14s \\ \hline
    \multicolumn{5}{|c|}{\textbf{Social Networks (SNAP)}} \\ \hline
    com-LiveJournal & 4.00M & 69.4M & 17.3 & 0.22s \\ \hline
    com-Orkut & 3.07M & 234M & 76.3 & 0.68s \\ \hline
    \multicolumn{5}{|c|}{\textbf{Road Networks (DIMACS10)}} \\ \hline
    asia\_osm & 12.0M & 25.4M & 2.1 & 0.10s \\ \hline
    europe\_osm & 50.9M & 108M & 2.1 & 0.35s \\ \hline
    \multicolumn{5}{|c|}{\textbf{Protein k-mer Graphs (GenBank)}} \\ \hline
    kmer\_A2a & 171M & 361M & 2.1 & 1.30s \\ \hline
    kmer\_V1r & 214M & 465M & 2.2 & 1.37s \\ \hline
  \bottomrule
  \end{tabular}
\end{table}

\subsection{Performance Comparison}
\label{sec:performance-comparison}  

\subsubsection{Loading a Graph from Disk}

We now compare the performance of PetGraph \cite{sverdrup2025petgraph}, SNAP \cite{leskovec2016snap}, SuiteSparse:GraphBLAS \cite{davis2023algorithm, davis2019algorithm}, cuGraph \cite{kang2023cugraph}, Aspen \cite{dhulipala2019low}, and our DiGraph in loading graphs from files into memory. PetGraph is a sequential Rust implementation, cuGraph is a parallel GPU-based implementation, and SNAP, SuiteSparse:GraphBLAS, Aspen, and our DiGraph are multicore implementations. For PetGraph, we use \texttt{MtxData::from\_file()} to read a Matrix Market (MTX) file, extracting the matrix shape, non-zero indices, and symmetry information. We initialize a directed graph (\texttt{DiGraphMap}) and add vertices corresponding to each matrix row. We then iterate over index pairs, adding the edges with a weight of $1$, inserting an additional reverse edge if the matrix is symmetric. We measure runtime using \texttt{Instant::now()} before and after loading. For SNAP, we first convert the MTX file to an EdgeList format, ensuring isolated vertices are included. We then load the EdgeList file as a \texttt{Graph} using \texttt{TSnap::LoadEdgeList()} and explicitly add any isolated vertices. We measure runtime using \texttt{clock()} before and after loading, excluding the conversion time from MTX to EdgeList. With SuiteSparse:GraphBLAS, we use \texttt{LAGraph\_MMRead()} to read the MTX file into a \texttt{GrB\_Matrix}, then construct a graph using the \texttt{LAGraph\_New()} function. We record the runtime using \texttt{LAGraph\_WallClockTime()}. Note that the LAGraph library is a collection of high-level graph algorithms built on the GraphBLAS C API, using linear algebra for graph computations.

For cuGraph, we first convert the MTX file to a CSV format. We then read the CSV into a dataframe using \texttt{cudf.read\_csv()}, setting edge weights to $1$. We create an empty graph $G$ with \texttt{cugraph.Gra} \texttt{ph()} and add edges using \texttt{G.from\_cudf\_edgelist()}. Runtime is measured using \texttt{time.time()}, before and after loading, and does not include the MTX-to-CSV conversion time. With Aspen, we first converting the MTX file to an Adjacency graph format, which is supported by Aspen, and is equivalent to the CSR format in plaintext. We then use \texttt{initialize\_graph()} to read the file into a versioned graph. Runtime is measured using \texttt{std::chrono::high\_resoluti} \texttt{on\_clock::now()} before and after loading, excluding the conversion time. However, as the Adjacency graph format is equivalent to CSR, this gives Aspen an unfair advantage. For our DiGraph, we use Algorithm \ref{alg:load} to load the MTX file into a CSR graph. We then measure runtime using \texttt{std::chrono::high\_resolution\_clock::now()} before and after execution of the algorithm. In all cases, runtime is averaged over five runs to minimize measurement noise.

\begin{figure*}[hbtp]
  \centering
  \subfigure[Runtime in seconds (logarithmic scale) with \textit{PetGraph}, \textit{SNAP}, \textit{SuiteSparse:GraphBLAS}, \textit{cuGraph}, \textit{Aspen}, and \textit{Our DiGraph}]{
    \label{fig:static-read--runtime}
    \includegraphics[width=0.98\linewidth]{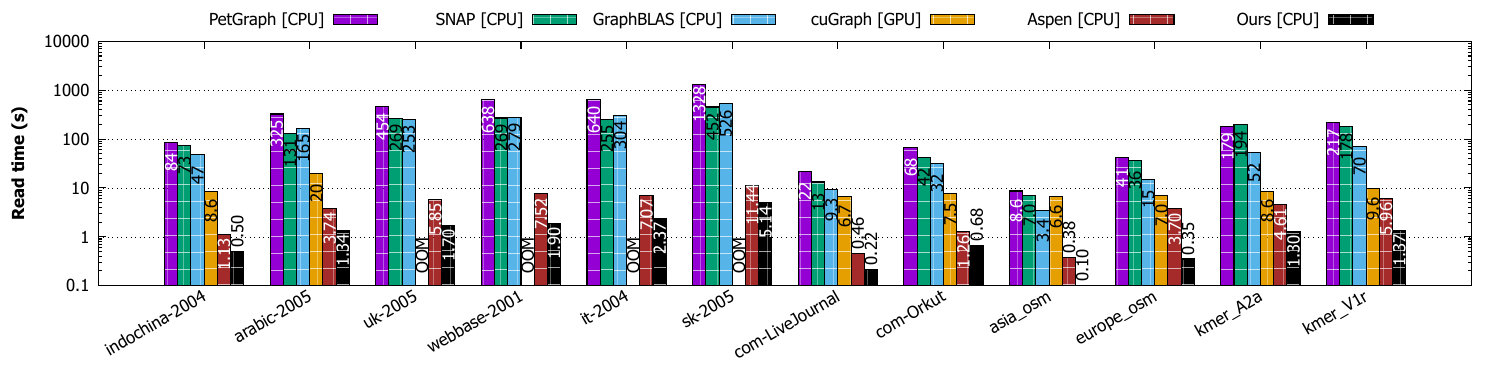}
  }
  \subfigure[Speedup of \textit{Our DiGraph} (logarithmic scale) with respect to \textit{PetGraph}, \textit{SNAP}, \textit{SuiteSparse:GraphBLAS}, \textit{cuGraph}, and \textit{Aspen}.]{
    \label{fig:static-read--speedup}
    \includegraphics[width=0.98\linewidth]{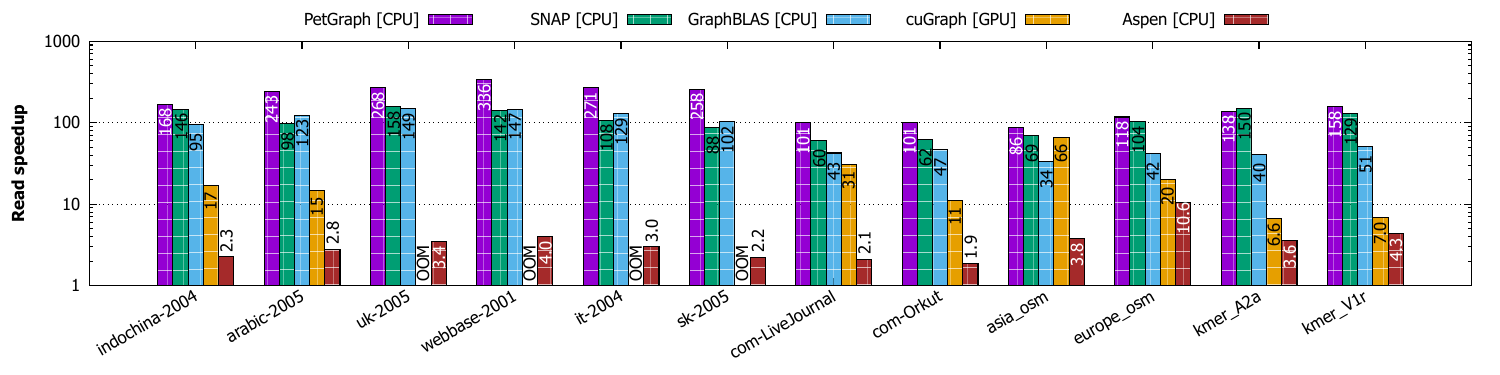}
  } \\[-2ex]
  \caption{Runtime in seconds (logarithmic scale) and speedup (logarithmic scale) for loading a graph from file into memory with \textit{PetGraph}, \textit{SNAP}, \textit{SuiteSparse:GraphBLAS}, \textit{cuGraph}, \textit{Aspen}, and \textit{Our DiGraph} for each graph in the dataset.}
  \label{fig:static-read}
\end{figure*}

Figure \ref{fig:static-read--runtime} presents the runtime for loading graphs from files into memory for each graph in the dataset (Table \ref{tab:dataset}) across PetGraph, SNAP, SuiteSparse:GraphBLAS, cuGraph, Aspen, and Our DiGraph, while Figure \ref{fig:static-read--speedup} shows the speedup of Our DiGraph relative to the other implementations. Due to out-of-memory issues, cuGraph fails to load the \textit{uk-2005}, \textit{webbase-2001}, \textit{it-2004}, and \textit{sk-2005} graphs, so these results are omitted. Our DiGraph achieves a mean speedup of $177\times$, $106\times$, $76\times$, $17\times$, and $3.3\times$ compared to PetGraph, SNAP, SuiteSparse:GraphBLAS, cuGraph, and Aspen, respectively. On the \textit{sk-2005} graph, it loads data in just $5.1$ seconds, reaching a graph loading rate of $379$ million edges per second. Notably, as discussed above, Aspen loads graphs\ignore{from disk} in the Adjacency graph format, equivalent to CSR, which likely gives it an unfair advantage over other frameworks. These results highlight the potential for significant improvements in graph loading times for state-of-the-art graph processing frameworks, and Algorithm \ref{alg:load} can help achieve this.

\subsubsection{Cloning a Graph}

Next, we compare the performance of PetGraph, SNAP, SuiteSparse:GraphBLAS, cuGraph, Aspen, and our DiGraph in cloning or obtaining a snapshot of a graph $G$. In PetGraph, cloning is performed using $G.clone()$. For SNAP, we find that using \texttt{TSnap::ConvertGraph()} is slower than reloading the graph from disk, so we opt for the latter. With SuiteSparse:GraphBLAS, we clone a graph by duplicating its adjacency matrix with \texttt{GrB\_Matrix\_dup()} and converting it to a graph via \texttt{LAGraph\_New()}. In cuGraph, we copy the dataframe containing the graph's edge list and convert it into a graph using \texttt{cugraph.Graph()} and $G'.from\_cudf\_edgelist()$, where $G'$ is the new graph. For Aspen, we take a snapshot of the graph with $G.acquire\_version()$. Finally, Our DiGraph uses Algorithm \ref{alg:clone} for cloning. Runtime is measured as discussed earlier, averaged over five runs.

\begin{figure*}[hbtp]
  \centering
  \subfigure[Runtime in seconds (logarithmic scale) with \textit{PetGraph}, \textit{SNAP}, \textit{SuiteSparse:GraphBLAS}, \textit{cuGraph}, \textit{Aspen}, and \textit{Our DiGraph}]{
    \label{fig:static-clone--runtime}
    \includegraphics[width=0.98\linewidth]{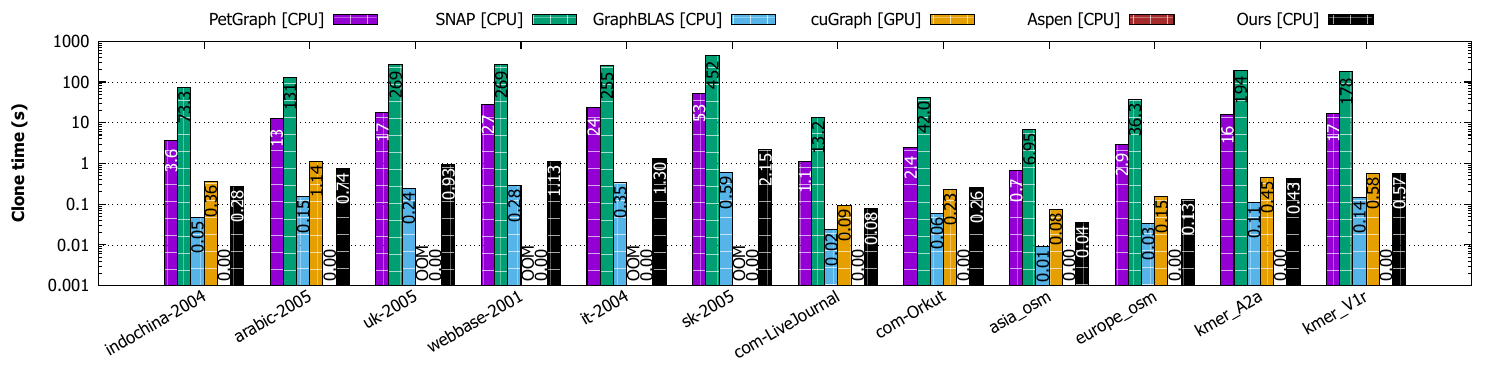}
  }
  \subfigure[Speedup of \textit{Our DiGraph} (logarithmic scale) with respect to \textit{PetGraph}, \textit{SNAP}, \textit{SuiteSparse:GraphBLAS}, \textit{cuGraph}, and \textit{Aspen}.]{
    \label{fig:static-clone--speedup}
    \includegraphics[width=0.98\linewidth]{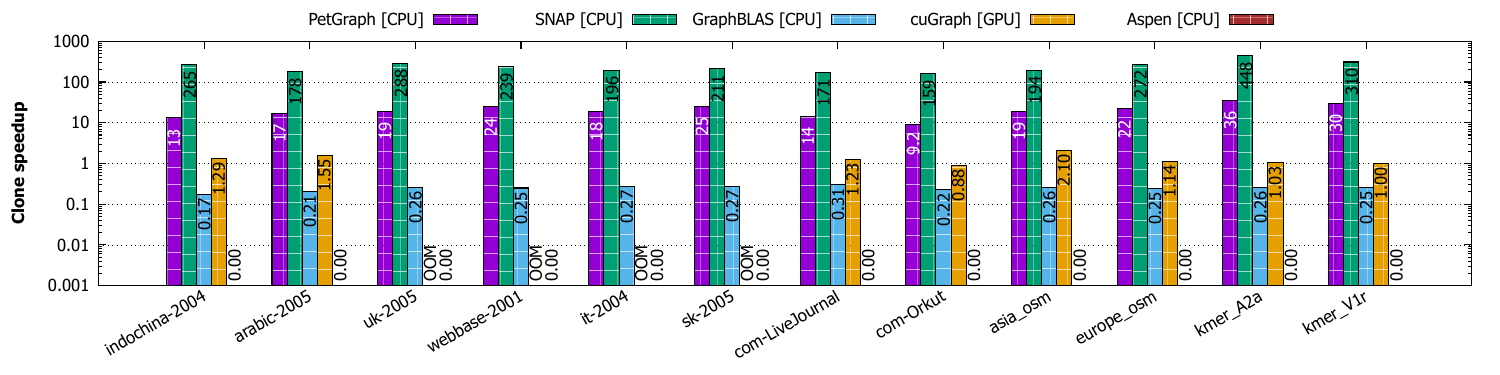}
  } \\[-2ex]
  \caption{Runtime in seconds (logarithmic scale) and speedup (logarithmic scale) for cloning a graph with \textit{PetGraph}, \textit{SNAP}, \textit{SuiteSparse:GraphBLAS}, \textit{cuGraph}, \textit{Aspen}, and \textit{Our DiGraph} for each graph in the dataset.}
  \label{fig:static-clone}
\end{figure*}

Figure \ref{fig:static-clone--runtime} presents the runtime for cloning graphs, for each graph in the dataset, across PetGraph, SNAP, SuiteSparse:Graph-BLAS, cuGraph, Aspen, and our DiGraph, while Figure \ref{fig:static-clone--speedup} shows the speedup of our DiGraph relative to the other implementations. Due to out-of-memory issues, results of cuGraph are omitted for the \textit{uk-2005}, \textit{webbase-2001}, \textit{it-2004}, and \textit{sk-2005} graphs. Our DiGraph achieves a mean speedup of $20\times$, $235\times$, $0.24\times$, $1.3\times$, and $0\times$ compared to PetGraph, SNAP, SuiteSparse:GraphBLAS, cuGraph, and Aspen, respectively. Aspen’s graph cloning is effectively zero-cost since we take a snapshot with its $G.acquire\_version()$, which only requires pointing to the root node of the vertex tree. In addition, SuiteSparse:GraphBLAS likely performs a shallow copy of the adjacency matrix, copying individual rows/columns only when modified. In contrast, PetGraph, SNAP, cuGraph, and our DiGraph perform full deep copies. On the \textit{sk-2005} graph, our DiGraph deep copies the graph in only $2.1$ seconds, thus achieving a rate of $908M$ edges/s. These results highlight that our DiGraph performs efficient deep copies (using Algorithm \ref{alg:clone}), while also demonstrating the significant advantages of Aspen’s C-trees and the lazy copy behavior of SuiteSparse:GraphBLAS.

\begin{figure*}[hbtp]
  \centering
  \subfigure[Runtime split for loading an MTX file into a CSR representation using Algorithm \ref{alg:load}.]{
    \label{fig:read-split-runtime}
    \includegraphics[width=0.48\linewidth]{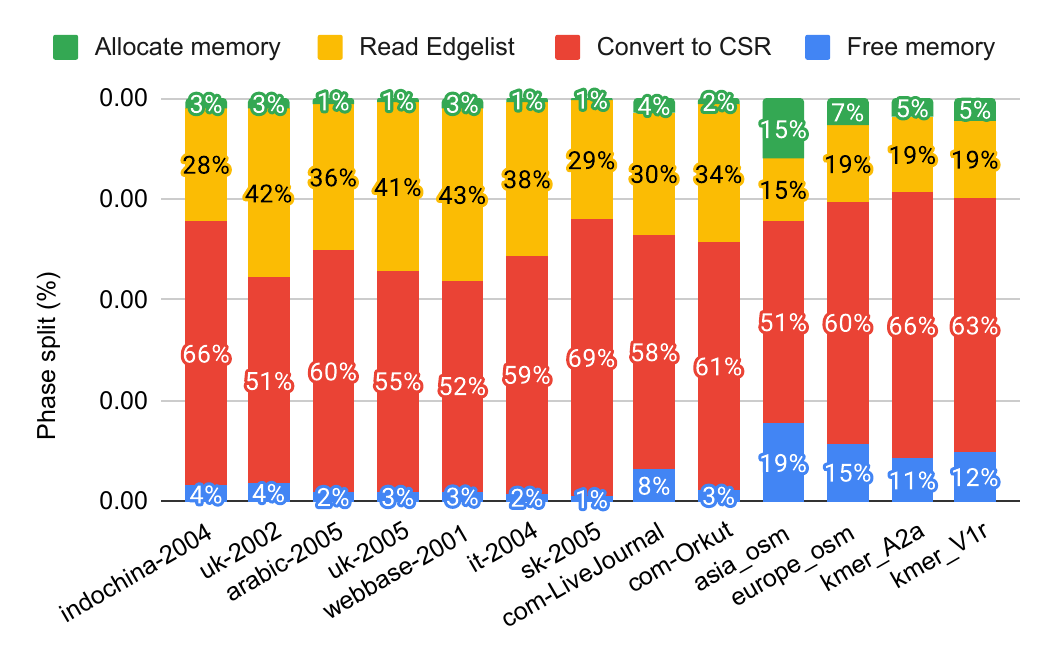}
  }
  \subfigure[Runtime split for cloning a CSR graph into \textit{Our DiGraph} using Algorithm \ref{alg:clone}.]{
    \label{fig:clone-split-runtime}
    \includegraphics[width=0.48\linewidth]{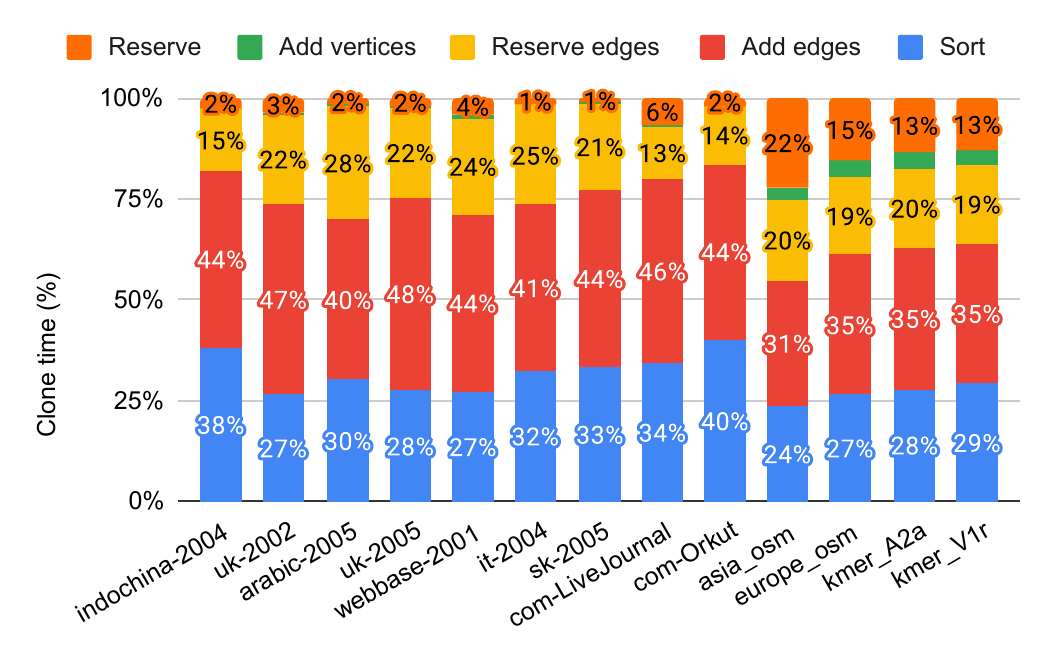}
  } \\[-2ex]
  \caption{Runtime split for loading a graph from an MTX file into a CSR representation using Algorithm \ref{alg:load} is shown on the left, while the runtime split for cloning a CSR graph into \textit{Our DiGraph} using Algorithm \ref{alg:clone} is on the right. The cloning of \textit{Our DiGraph} into another one has a similar runtime split.}
  \label{fig:split-runtime}
\end{figure*}

\subsubsection{Performing Edge Deletions}
\label{sec:perform-edge-deletions}

We now evaluate performance of the graph implementations --- PetGraph, SNAP, SuiteSparse:Graph-BLAS, cuGraph, Aspen, and our DiGraph --- in performing a batch of edge deletions, both in-place and by creating a new graph instance without the specified edges. For PetGraph, edge deletions are performed using the method $G.remove\_edge()$ in a loop over the batch of edges to be deleted. To perform edge deletions onto a new graph instance, we simply clone the original graph and remove the edges using the same method. In SNAP, edges are removed by using $G.DelEdge()$ iteratively, while for new graph instances, we reload the graph from disk and remove the edges accordingly. With SuiteSparse:GraphBLAS, we remove edges by deleting entries from its adjacency matrix via \texttt{GrB\_Matrix\_removeElement()}. For a new graph instance, we duplicate the adjacency matrix with \texttt{GrB\_Matrix\_dup()}, create a new graph using \texttt{LAGraph\_New()}, and remove edges in the same way. In cuGraph, we construct a new edge list through a left merge on the original, followed by drop and rename operations, then build a new graph $G'$ using $cugraph.Graph()$ and $G'.from\_cudf\_edgelist()$. For new-instance deletions, the original edge list is copied before applying the same process. For Aspen, we delete edges by calling $G.delete\_edges\_batch()$, which returns a new graph snapshot with the specified edges removed. There is no in-place deletion mechanism in Aspen. Finally, our DiGraph removes edges using \texttt{subtractGraphInplace()} in Algorithm \ref{alg:sub}. To perform edge deletions on a new graph instance, we use the \texttt{subtractGraph()} function instead, which is more efficient than cloning the original graph and then removing the edges in-place.

\begin{figure*}[hbtp]
  \centering
  \subfigure[Overall result]{
    \label{fig:dynamic-del-runtime--mean}
    \includegraphics[width=0.38\linewidth]{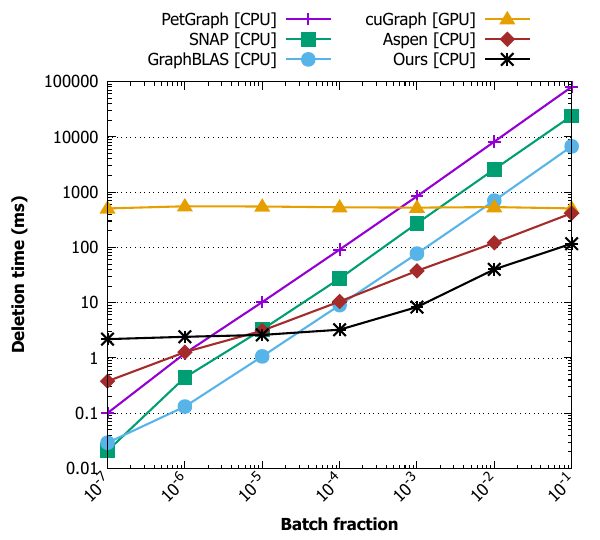}
  }
  \subfigure[Results on each graph]{
    \label{fig:dynamic-del-runtime--all}
    \includegraphics[width=0.58\linewidth]{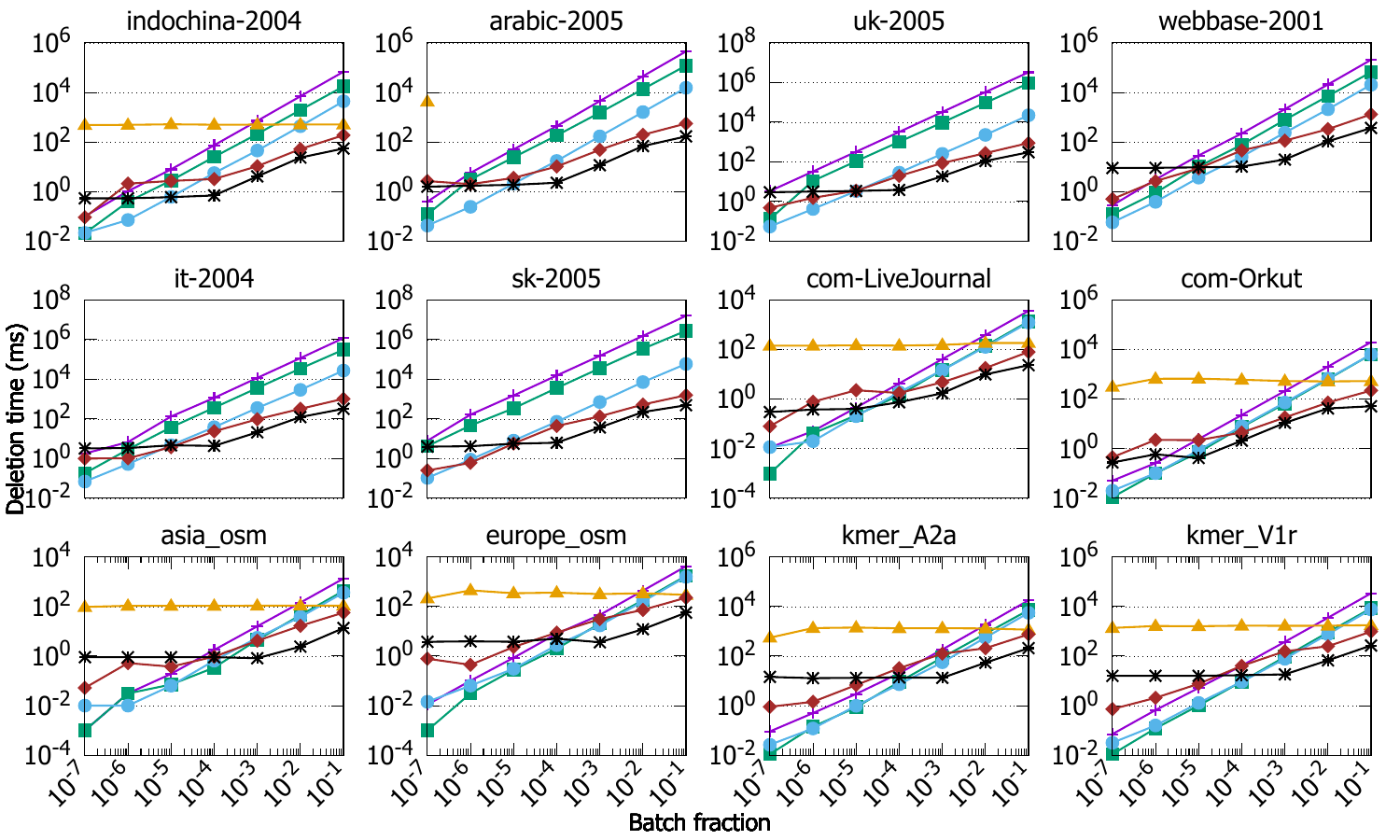}
  } \\[-2ex]
  \caption{Runtime in milliseconds (logarithmic scale) of \textit{deleting} a batch of $10^{-7}|E|$ to $0.1|E|$ randomly generated edges into a graph (\textit{in-place}), in multiples of $10$. Here, we evaluate \textit{PetGraph}, \textit{SNAP}, \textit{SuiteSparse:GraphBLAS}, \textit{cuGraph}, \textit{Aspen}, and \textit{Our DiGraph} on each graph in the dataset. The left subfigure presents overall runtimes using the geometric mean for consistent scaling, while the right subfigure shows runtimes for individual graphs.}
  \label{fig:dynamic-del-runtime}
\end{figure*}

\begin{figure*}[hbtp]
  \centering
  \subfigure[Overall result]{
    \label{fig:dynamic-delnew-runtime--mean}
    \includegraphics[width=0.38\linewidth]{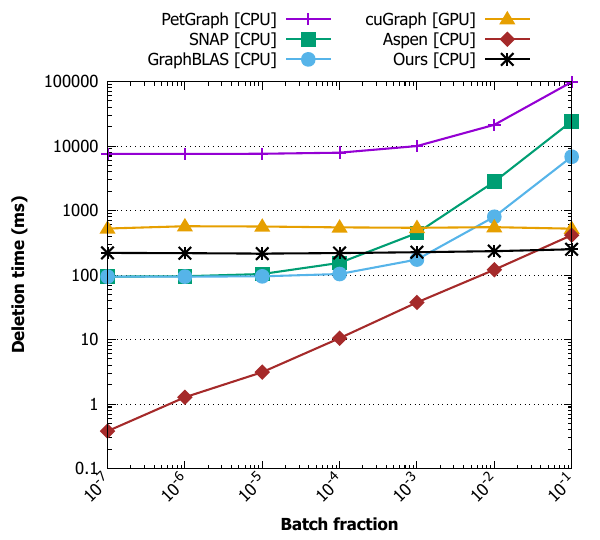}
  }
  \subfigure[Results on each graph]{
    \label{fig:dynamic-delnew-runtime--all}
    \includegraphics[width=0.58\linewidth]{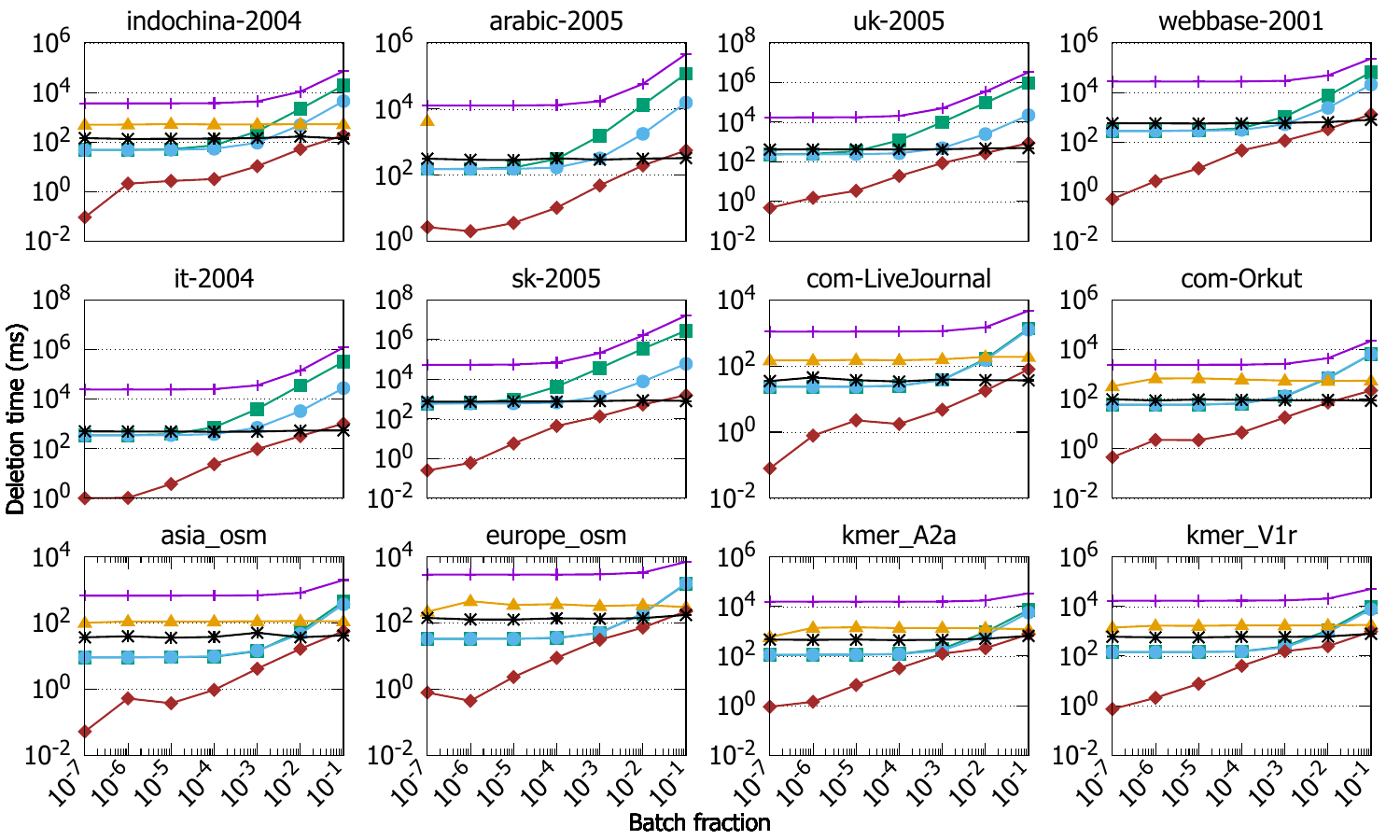}
  } \\[-2ex]
  \caption{Runtime in milliseconds (logarithmic scale) of \textit{deleting} a batch of $10^{-7}|E|$ to $0.1|E|$ randomly generated edges into a new graph, in multiples of $10$. Here, we evaluate \textit{PetGraph}, \textit{SNAP}, \textit{SuiteSparse:GraphBLAS}, \textit{cuGraph}, \textit{Aspen}, and \textit{Our DiGraph} on each graph in the dataset. The left subfigure presents overall runtimes using the geometric mean for consistent scaling, while the right subfigure shows runtimes for individual graphs.}
  \label{fig:dynamic-delnew-runtime}
\end{figure*}

We test the performance of these implementations\ignore{(in removing a subset of edges, both in-place and by creating a new graph instance without the specified edges)} on large graphs, from Table \ref{tab:dataset}, with random batch updates, consisting of edge deletions. Edges are uniformly deleted, with batch sizes ranging from $10^{-7} |E|$ to $0.1|E|$ (batch size is measured as a fraction of edges in the original graph), and multiple random batch updates per batch size are generated for averaging. Figure \ref{fig:dynamic-del-runtime--mean} shows the overall runtime of in-place deletions at each batch size across all implementations, while Figure \ref{fig:dynamic-del-runtime--all} details runtimes for each framework and batch size on individual graphs. Due to out-of-memory issues, results for cuGraph are omitted for \textit{uk-2005}, \textit{webbase-2001}, \textit{it-2004}, and the \textit{sk-2005} graphs. Additionally, Figure \ref{fig:dynamic-delnew-runtime--mean} illustrates the overall runtimes for each framework when deleting edges into a new graph instance, with Figure \ref{fig:dynamic-delnew-runtime--all} providing per-graph results.

Figure \ref{fig:dynamic-del-runtime--mean} shows that, on batch updates of size $10^{-4}|E|$ to $0.1|E|$, our DiGraph offers a mean speedup of $141\times$, $44\times$, $13\times$, $28\times$, and $3.5\times$ over PetGraph, SNAP, SuiteSparse:GraphBLAS, cuGraph, and Aspen, respectively, for in-place edge deletions. This is due to efficient parallel deletion of edges using $\textit{setDifference()}$ upon independent vertices (in Algorithms \ref{alg:digraph2} and \ref{alg:sub}). However, for small batch updates, the overhead of iterating over all vertices, leads to lower performance. When deleting edges into a new graph instance, our DiGraph achieves a mean speedup of $87\times$, $6.3\times$, $2.4\times$, $2.3\times$, and $0.29\times$ over PetGraph, SNAP, SuiteSparse:GraphBLAS, cuGraph, and Aspen, respectively. These results underscore the limitations of current state-of-the-art graph processing frameworks in handling large batch updates (both in-place and when creating a new graph instance) and highlight the advantages of C-trees, as used by Aspen, which enable lightweight graph snapshots and superior performance in batch updates involving new graph instances. They also suggest that our DiGraph could benefit from further optimizations, particularly for small batch sizes.

\subsubsection{Performing Edge Insertions}

Next, we evaluate the performance of the graph implementations in performing a batch of edge insertions, both in-place and by creating a new graph instance that includes the newly added edges. For PetGraph, edges are inserted using the method $G.add\_edge()$ in a loop over the batch of edges to be inserted. To insert edges into a new graph instance, we first clone the original graph and then add the edges using the same method. In SNAP, edges are inserted iteratively using $G.AddEdge()$, while for new graph instances, we reload the graph from disk and then perform insertions. In SuiteSparse:GraphBLAS, edges are inserted by updating entries in its adjacency matrix using \texttt{GrB\_Matrix\_setElement()}. For a new graph instance, we duplicate the adjacency matrix with \texttt{GrB\_Matrix\_dup()}, create a new graph using \texttt{LAGraph\_New()}, and then insert edges using the same method. In cuGraph, we construct a new edge list by performing a left merge operation with the original, dropping the duplicate entries, renaming the column name in the dataframe, and concatenating the insertions, before building a new graph $G'$ using $cugraph.Graph()$ and $G'.from\_cudf\_edgelist()$. For new-instance insertions, the original edge list is copied before undergoing the same process. In Aspen, edges are inserted using $G.insert\_edges\_b$ $atch()$, which produces a new graph snapshot containing the added edges, as Aspen does not support in-place insertions. Finally, our DiGraph inserts edges using \texttt{addGraphInplace()} as described in Algorithm \ref{alg:add}. To perform insertions into a new graph instance, we use \texttt{addGraph()}, which is more efficient than cloning the original graph and then inserting edges in-place.

\begin{figure*}[hbtp]
  \centering
  \subfigure[Overall result]{
    \label{fig:dynamic-ins-runtime--mean}
    \includegraphics[width=0.38\linewidth]{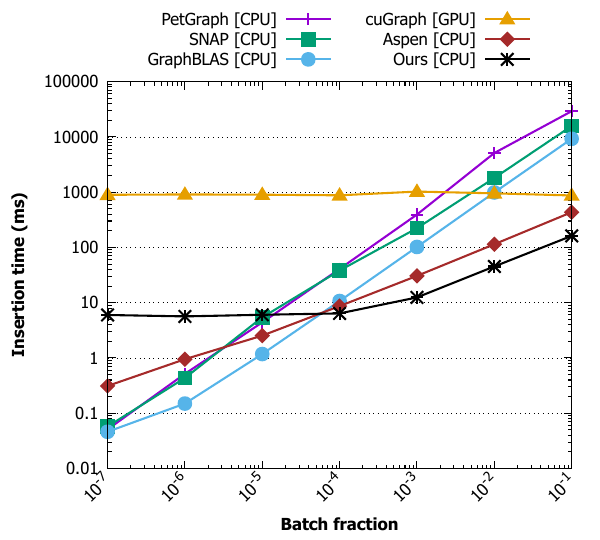}
  }
  \subfigure[Results on each graph]{
    \label{fig:dynamic-ins-runtime--all}
    \includegraphics[width=0.58\linewidth]{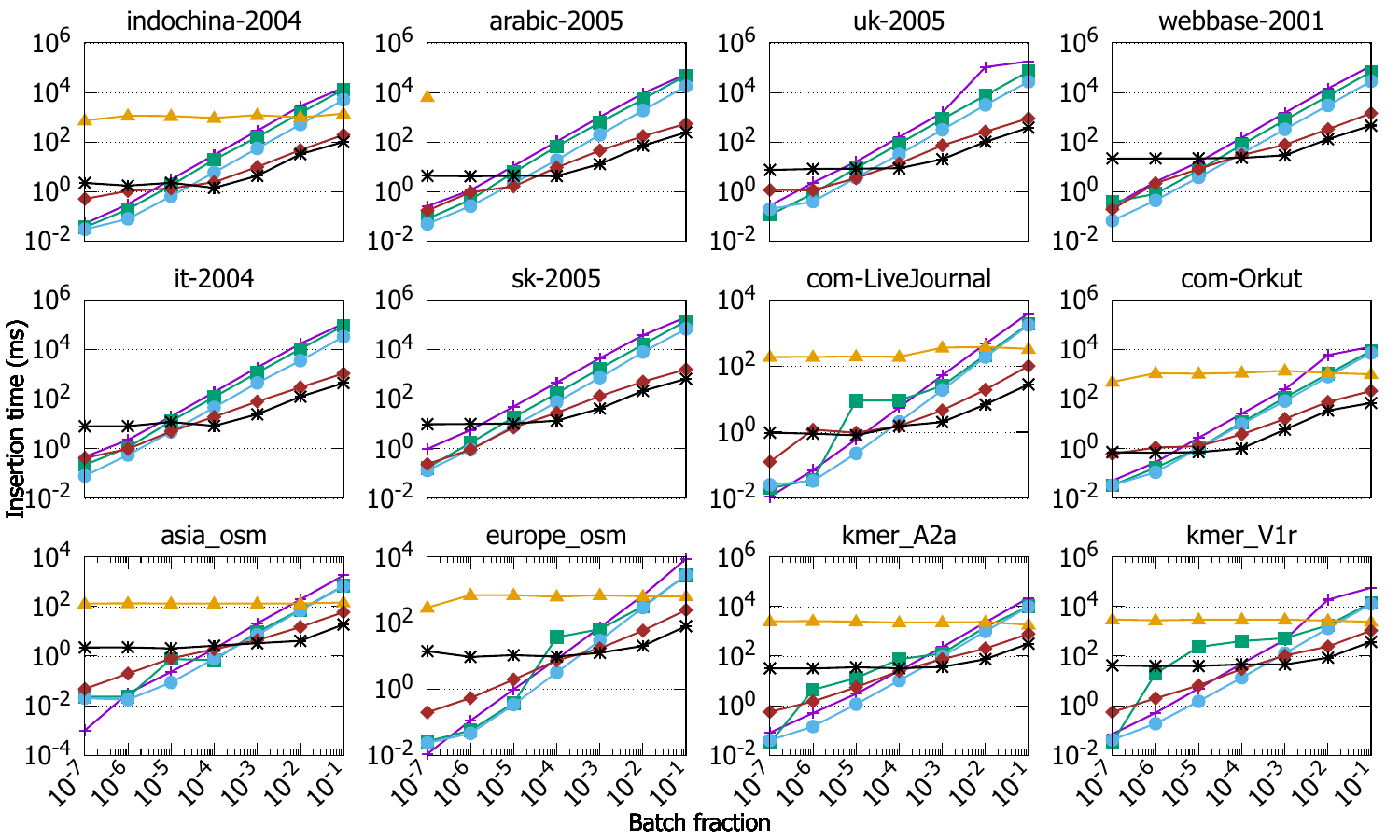}
  } \\[-2ex]
  \caption{Runtime in milliseconds (logarithmic scale) of \textit{inserting} a batch of $10^{-7}|E|$ to $0.1|E|$ randomly generated edges into a graph (\textit{in-place}), in multiples of $10$. Here, we evaluate \textit{PetGraph}, \textit{SNAP}, \textit{SuiteSparse:GraphBLAS}, \textit{cuGraph}, \textit{Aspen}, and \textit{Our DiGraph} on each graph in the dataset. The left subfigure presents overall runtimes using the geometric mean for consistent scaling, while the right subfigure shows runtimes for individual graphs.}
  \label{fig:dynamic-ins-runtime}
\end{figure*}

\begin{figure*}[hbtp]
  \centering
  \subfigure[Overall result]{
    \label{fig:dynamic-insnew-runtime--mean}
    \includegraphics[width=0.38\linewidth]{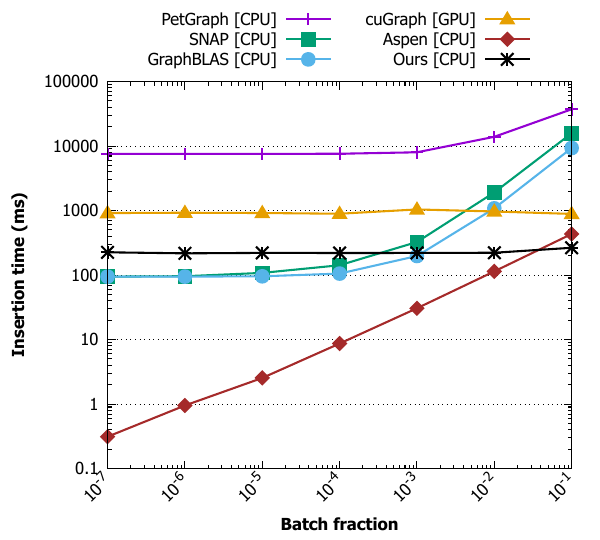}
  }
  \subfigure[Results on each graph]{
    \label{fig:dynamic-insnew-runtime--all}
    \includegraphics[width=0.58\linewidth]{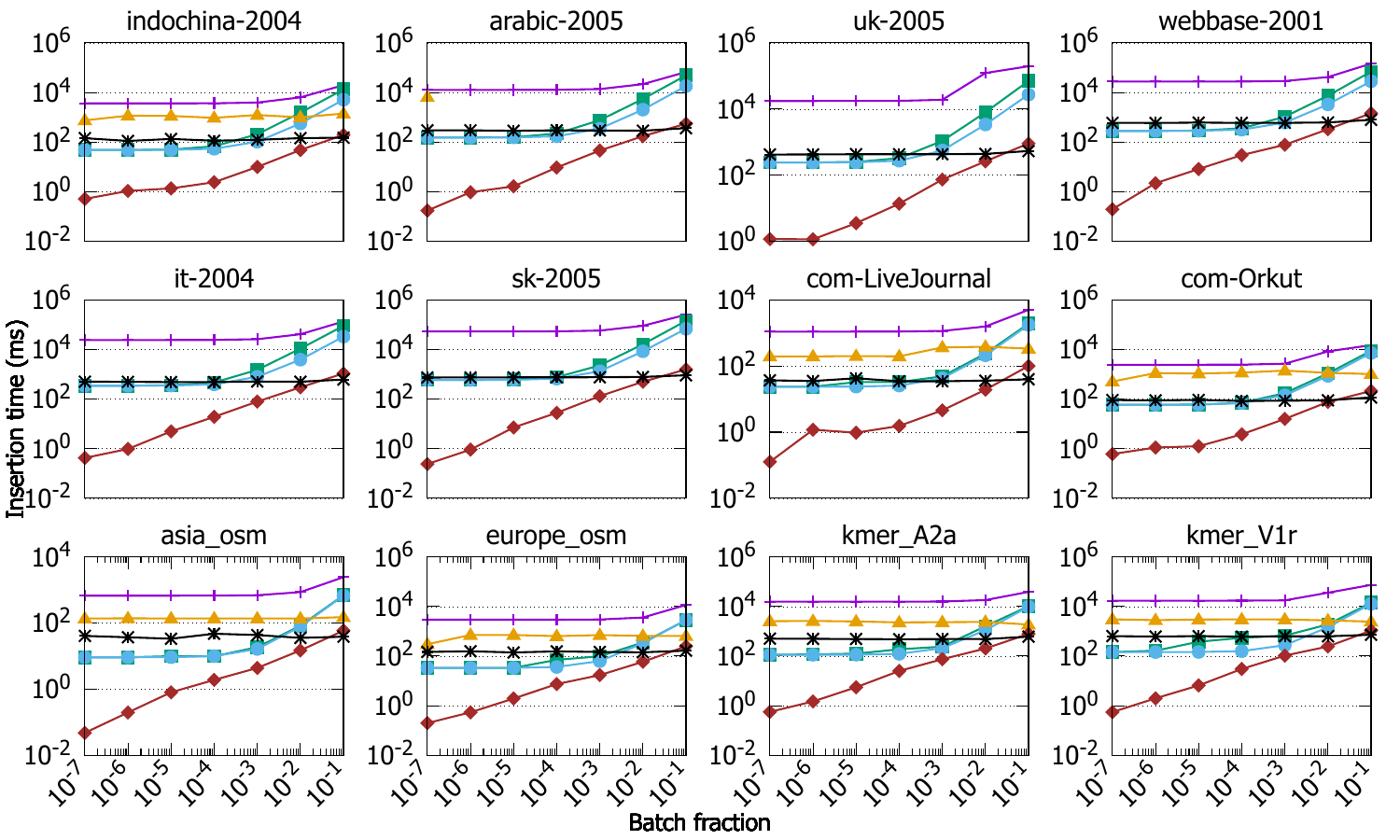}
  } \\[-2ex]
  \caption{Runtime in milliseconds (logarithmic scale) of \textit{inserting} a batch of $10^{-7}|E|$ to $0.1|E|$ randomly generated edges into a new graph, in multiples of $10$. Here, we evaluate \textit{PetGraph}, \textit{SNAP}, \textit{SuiteSparse:GraphBLAS}, \textit{cuGraph}, \textit{Aspen}, and \textit{Our DiGraph} on each graph in the dataset. The left subfigure presents overall runtimes using the geometric mean for consistent scaling, while the right subfigure shows runtimes for individual graphs.}
  \label{fig:dynamic-insnew-runtime}
\end{figure*}

We test the performance of these implementations on large graphs from Table \ref{tab:dataset}, with random batch updates consisting of edge insertions. To prepare the set of edges for insertion, we select vertex pairs with equal probability, with batch sizes ranging from $10^{-7} |E|$ to $0.1|E|$ (batch size measured as a fraction of edges in the original graph). Multiple random batch updates per batch size are generated for averaging, as earlier. Figure \ref{fig:dynamic-ins-runtime--mean} presents the overall runtime of in-place insertions at each batch size across all implementations, while Figure \ref{fig:dynamic-ins-runtime--all} details runtimes for each framework and batch size on individual graphs.\ignore{Results for cuGraph are omitted for \textit{uk-2005}, \textit{webbase-2001}, \textit{it-2004}, and \textit{sk-2005} due to out-of-memory issues.} Additionally, Figure \ref{fig:dynamic-insnew-runtime--mean} illustrates the overall runtimes when inserting edges into a new graph instance, with Figure \ref{fig:dynamic-insnew-runtime--all} providing per-graph results.

Figure \ref{fig:dynamic-ins-runtime--mean} shows that for batch updates ranging from $10^{-4}|E|$ to $0.1|E|$, our DiGraph achieves a mean speedup of $45\times$, $25\times$, $11\times$, $34\times$, and $2.2\times$ over PetGraph, SNAP, SuiteSparse:GraphBLAS, cuGraph, and Aspen, respectively, for in-place edge insertions. This performance gain is due to efficient parallel insertion using $\textit{setUnion()}$ on independent vertices, as described in Algorithms \ref{alg:digraph2} and \ref{alg:add}. However, for small batch sizes, the overhead of iterating over all vertices reduces performance (as with edge deletions). When inserting edges into a new graph instance, our DiGraph attains a mean speedup of $58\times$, $4.7\times$, $2.9\times$, $4.1\times$, and $0.26\times$ over the same frameworks. These findings align with the conclusions discussed in Section \ref{sec:perform-edge-deletions}.

\subsubsection{Performance on $k$-step Reverse Walks}
\label{sec:perform-reverse-walks}

Finally, we evaluate the performance of various graph representations on a representative algorithm by measuring the efficiency of $k$-step reverse walks from each vertex in the given input graph, and counting the number of walks ending at each vertex. This corresponds to computing $A_T^k \cdot \vec{1}$, where $A_T$ is the transposed adjacency matrix and $\vec{1}$ is a ones vector. Reverse walks are preferred as they can be executed directly on the input graph, whereas forward walks require its transpose. The purpose of this evaluation is to assess the efficiency of different frameworks in processing graph algorithms on updated graphs, ensuring they prioritize algorithm execution alongside fast data structure updates. For PetGraph, SNAP, Aspen, and our DiGraph, we implement this using their respective APIs, following a similar approach to Algorithm \ref{alg:visit}. In SuiteSparse:GraphBLAS, we use the \texttt{GrB\_vxm()} function repeatedly to compute $A_T^k \cdot \vec{1}$, while in cuGraph, we utilize \texttt{cupyx.scatter\_add()}. Note however that PetGraph is executed sequentially, while the others are parallelized.

\begin{figure*}[hbtp]
  \centering
  \subfigure[Overall result]{
    \label{fig:dynamic-visdel-runtime--mean}
    \includegraphics[width=0.38\linewidth]{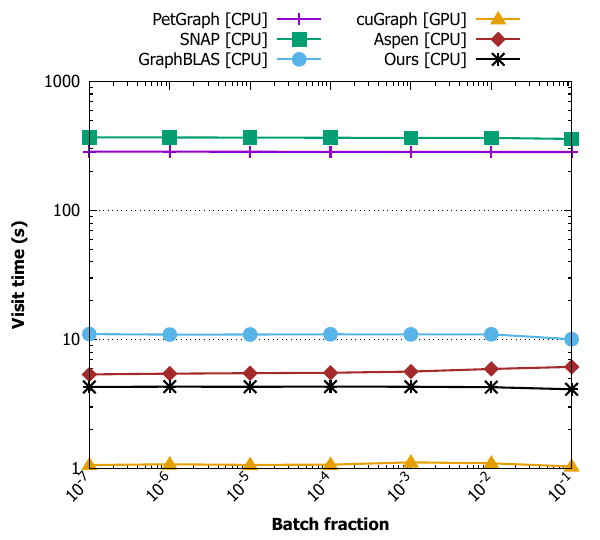}
  }
  \subfigure[Results on each graph]{
    \label{fig:dynamic-visdel-runtime--all}
    \includegraphics[width=0.58\linewidth]{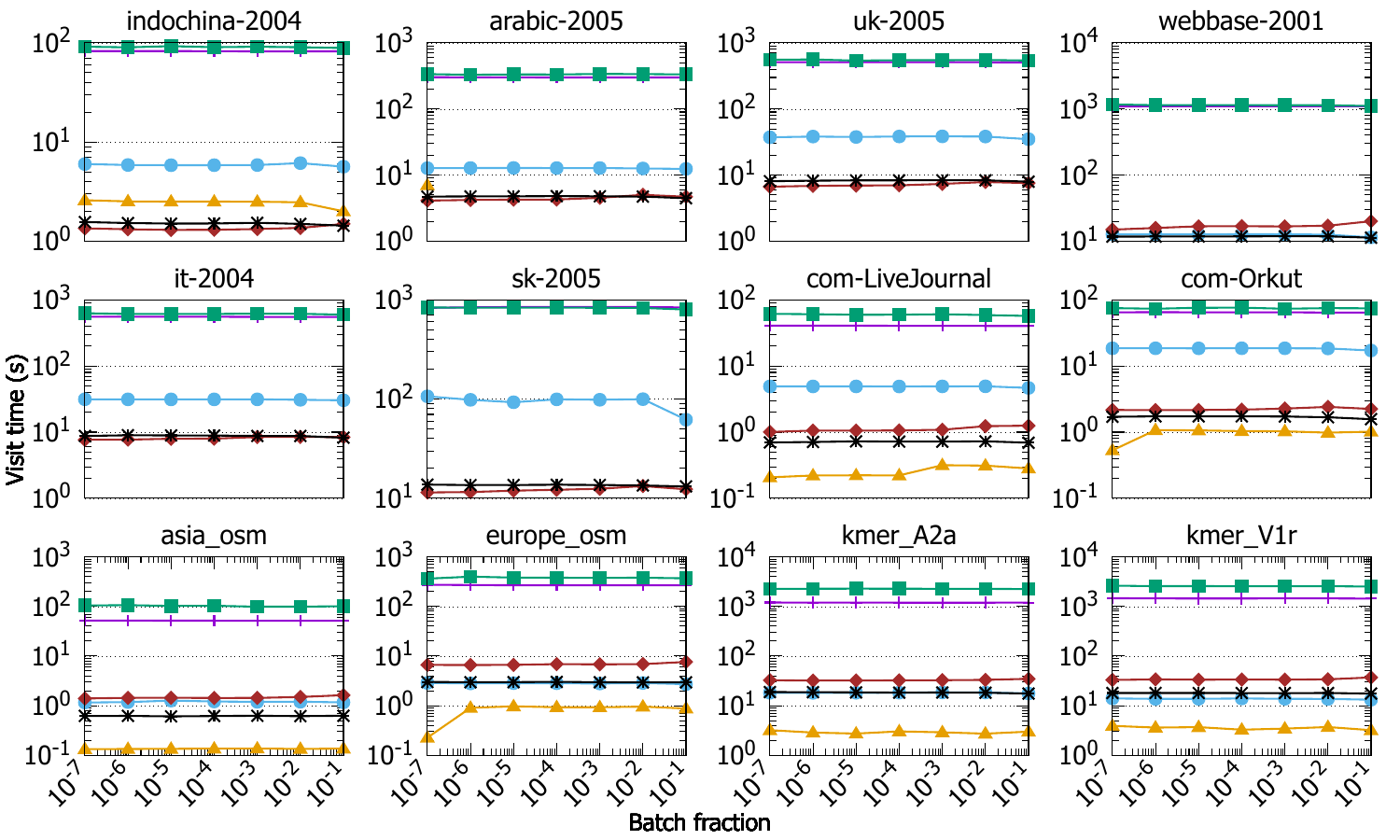}
  } \\[-2ex]
  \caption{Runtime in seconds (logarithmic scale) of performing $42$-step reverse walks from each vertex in a graph with randomly generated \textit{edge deletions} ranging from $10^{-7}|E|$ to $0.1|E|$ in multiples of $10$. Here, we evaluate \textit{PetGraph}, \textit{SNAP}, \textit{SuiteSparse:GraphBLAS}, \textit{cuGraph}, \textit{Aspen}, and \textit{Our DiGraph} on large graphs from Table \ref{tab:dataset}. The left subfigure presents overall runtimes using the geometric mean for consistent scaling, while the right subfigure shows visit times for individual graphs.}
  \label{fig:dynamic-visdel-runtime}
\end{figure*}

\begin{figure*}[hbtp]
  \centering
  \subfigure[Overall result]{
    \label{fig:dynamic-visins-runtime--mean}
    \includegraphics[width=0.38\linewidth]{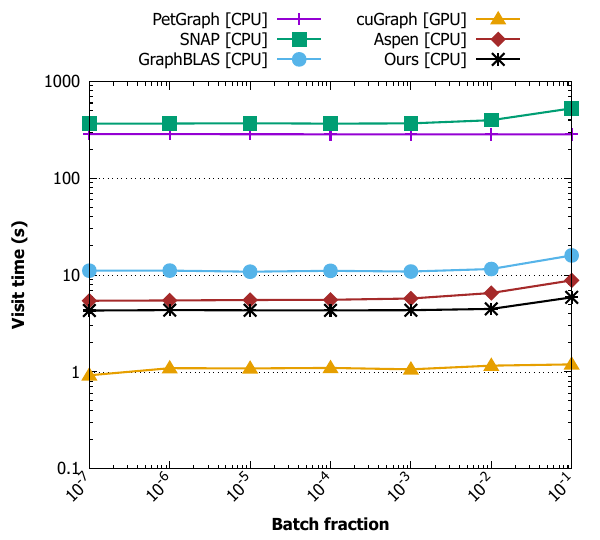}
  }
  \subfigure[Results on each graph]{
    \label{fig:dynamic-visins-runtime--all}
    \includegraphics[width=0.58\linewidth]{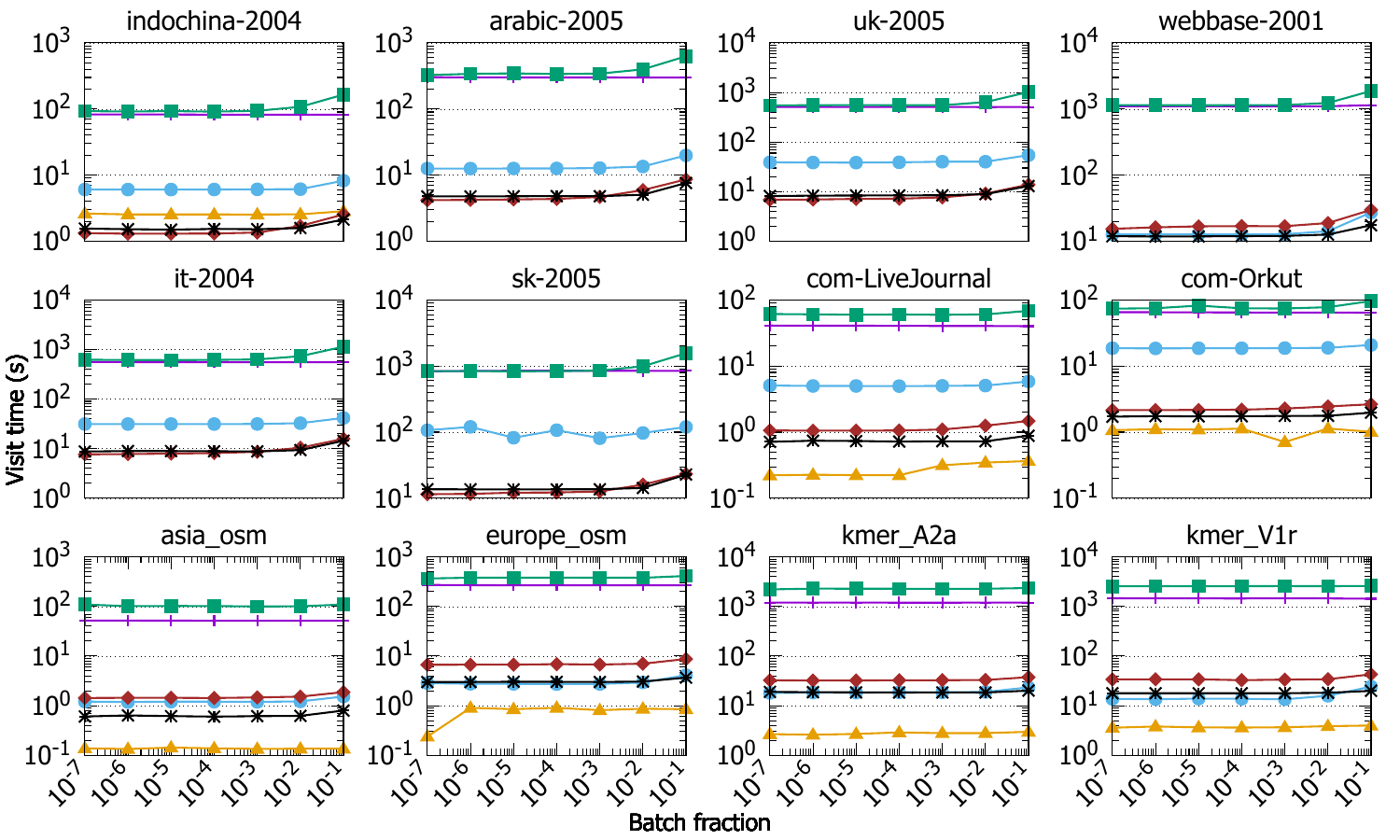}
  } \\[-2ex]
  \caption{Runtime (logarithmic scale) of performing $42$-step reverse walks from each vertex in a graph with randomly generated \textit{edge insertions} ranging from $10^{-7}|E|$ to $0.1|E|$ in multiples of $10$. Here, we evaluate \textit{PetGraph}, \textit{SNAP}, \textit{SuiteSparse:GraphBLAS}, \textit{cuGraph}, \textit{Aspen}, and \textit{Our DiGraph} on large graphs from Table \ref{tab:dataset}. The left subfigure presents overall runtimes using the geometric mean for consistent scaling, while the right subfigure shows visit times for individual graphs.}
  \label{fig:dynamic-visins-runtime}
\end{figure*}

We evaluate the efficiency of the updated graphs in each graph processing framework by testing their performance on the dataset, using random batch updates consisting of either edge deletions or insertions. Edges are uniformly deleted, while insertion edges are selected by choosing vertex pairs with equal probability. Batch sizes range from $10^{-7} |E|$ to $0.1|E|$, as earlier, and multiple random batch updates per batch size are generated for averaging. Figure \ref{fig:dynamic-visdel-runtime--mean} presents the overall runtime of performing $42$-step reverse walks on graphs with edge deletions at each batch size across all implementations, while Figure \ref{fig:dynamic-visdel-runtime--all} details runtimes per framework and graph. Due to out-of-memory issues, cuGraph results are omitted for \textit{uk-2005}, \textit{webbase-2001}, \textit{it-2004}, and \textit{sk-2005}. Similarly, Figure \ref{fig:dynamic-visins-runtime--mean} shows overall runtimes for $42$-step reverse walks with edge insertions, with Figure \ref{fig:dynamic-visins-runtime--all} providing per-graph results.

Figure \ref{fig:dynamic-visdel-runtime--mean} illustrates that for batch updates ranging from $10^{-7}|E|$ to $0.1|E|$, our DiGraph achieves an average speedup of $67\times$, $86\times$, $2.5\times$, $0.25\times$, and $1.3\times$ over PetGraph, SNAP, SuiteSparse:GraphBLAS, cuGraph, and Aspen, respectively, when performing $k$-step reverse walks on graphs with edge deletions. This performance gain is largely due to the fact that all edges are stored contiguously, even after a batch update, and our DiGraph is designed in a Struct-of-Arrays (SoA) format, both of which enable high cache locality. cuGraph outperforms our DiGraph because it is being executed on the GPU. We also note that the performance of Aspen starts to slightly degrade as the batch size increases, which is likely due to fragmentation in the C-tree. When executing reverse walks on graphs with newly inserted edges, our DiGraph achieves mean speedups of $63\times$, $86\times$, $2.6\times$, $0.24\times$, and $1.3\times$ over the same frameworks. These results point to the inefficiencies in representations of existing graph processing frameworks, but also highlight the advantages of C-trees, as utilized by Aspen, which offer performance similar to our DiGraph, with the added advantage of fast graph snapshots.

\section{Conclusion}
\label{sec:conclusion}
In this report, we evaluated the performance of several state-of-the-art dynamic graph processing frameworks and demonstrated that our custom implementation, DiGraph --- which uses our CP2AA allocator, contiguous edge arrays, and a Struct-of-Arrays (SoA) approach --- achieves significant performance gains across key tasks, including graph loading, cloning, in-place dynamic graph updates, and graph algorithm execution (iterative traversal). Our results also highlight the benefits of lazy copying (SuiteSparse:GraphBLAS), and zero-cost snapshotting (Aspen) for applications which need to preserve the original graph, making in-place graph updates unsuitable. Future work could optimize for small batch updates.

\begin{acks}
I would like to thank Prof. Kishore Kothapalli, Prof. Sathya Peri, and Prof. Dip Sankar Banerjee for their support.
\end{acks}

\bibliographystyle{ACM-Reference-Format}
\bibliography{main}

\clearpage
\clearpage
\appendix
\section{Appendix}

\subsection{Memory Allocators}
\label{sec:memory-allocators}

We now discuss a few memory allocators, which support the design of our graph representation.

\begin{algorithm}[H] 
\caption{A Fixed-capacity Arena Allocator (FAA).}
\label{alg:faa}
\begin{algorithmic}[1]
\Require{$\textsc{alloc\_size}_a$: Size of each allocation (constant)}
\Require{$\textsc{pool\_size}_a$: Size of the memory block/pool (constant)}
\Require{$pool_a$: A memory block for the pool}

\Statex

\State \textbf{struct} $FAA \langle \textsc{alloc\_size}_a, \textsc{pool\_size}_a \rangle (pool_a)$
\State \ \ $\textsc{alloc\_size} \gets \textsc{alloc\_size}_a$ \Comment{Size of each allocation} \label{alg:faa--init-const-begin}
\State \ \ $\textsc{pool\_size} \gets \textsc{pool\_size}_a$ \Comment{Size of the memory pool} \label{alg:faa--init-const-end}
\State \ \ $freed \gets \{\}$ \Comment{Freed allocations, can be reused} \label{alg:faa--init-begin}
\State \ \ $used \gets 0$ \Comment{Bytes used in the memory pool}
\State \ \ $pool \gets pool_a$ \Comment{The memory pool} \label{alg:faa--init-end}

\Statex

\State $\rhd$ Allocate memory, of size $A.\textsc{alloc\_size}$
\Function{allocate}{$ $} \textbf{of} FAA
  \State $\rhd$ Allocate from freed list, if available
  \If{$freed \neq \{\}$} \ReturnInline{$freed.pop()$} \label{alg:faa--allocate-freed}
  \EndIf
  \State $\rhd$ Allocate from pool
  \If{$used < \textsc{pool\_size}$} \label{alg:faa--allocate-pool-begin}
    \State $ptr \gets pool + used$
    \State $used \gets used + \textsc{alloc\_size}$
    \Return{$ptr$}
  \EndIf \label{alg:faa--allocate-pool-end}
  \Return{$\phi$} \label{alg:faa--allocate-phi}
\EndFunction

\Statex

\State $\rhd$ Free allocated memory
\Function{deallocate}{$ptr$} \textbf{of} FAA
  \State $freed.push(ptr)$ \label{alg:faa--deallocate-push}
\EndFunction

\Statex

\State $\rhd$ Free all allocated memory
\Function{reset}{$ $} \textbf{of} FAA
  \State $freed \gets \{\}$ \label{alg:faa--reset-begin}
  \State $used \gets 0$ \label{alg:faa--reset-end}
\EndFunction
\end{algorithmic}
\end{algorithm}

\subsubsection{Fixed Arena Allocator (FAA)}
\label{sec:faa}

We now discuss the Fixed-capacity Arena Allocator (FAA), outlined in Algorithm \ref{alg:faa}, which provides a lightweight, efficient memory allocation strategy for scenarios requiring frequent allocations and deallocations within a fixed memory budget. FAA minimizes fragmentation by recycling memory blocks and allows for rapid reset operations, making it well-suited for high-performance applications. It operates through three primary functions: \texttt{allocate()}, \texttt{deallocate()}, and \texttt{reset()}.

The allocator is instantiated with a predefined allocation size $\textsc{alloc\_size}_a$ and a total pool size $\textsc{pool\_size}_a$ (lines \ref{alg:faa--init-const-begin}-\ref{alg:faa--init-const-end}). The allocator maintains an internal memory pool $pool$, a counter $used$ tracking the number of allocated bytes, and a list $freed$ to store deallocated memory chunks for reuse (lines \ref{alg:faa--init-begin}-\ref{alg:faa--init-end}). Memory allocation is handled by the \texttt{allocate()} function. First, if any previously freed memory blocks exist in $freed$, one is retrieved and returned (line \ref{alg:faa--allocate-freed}). Otherwise, allocation proceeds from the main memory pool. If there is available space in the pool (i.e., $used < \textsc{pool\_size}$), the function assigns a pointer to the next available block, updates the $used$ counter, and returns the pointer (lines \ref{alg:faa--allocate-pool-begin}-\ref{alg:faa--allocate-pool-end}). If the pool is exhausted, the function returns a null pointer $\phi$, indicating allocation failure (line \ref{alg:faa--allocate-phi}). The \texttt{deallocate()} function allows memory to be freed by pushing the given pointer into the $freed$ list, making it available for future reuse (line \ref{alg:faa--deallocate-push}). To completely reset the allocator, the \texttt{reset()} function clears the $freed$ list and resets the $used$ counter to zero, effectively deallocating everything (lines \ref{alg:faa--reset-begin}-\ref{alg:faa--reset-end}).

\subsubsection{Variable-capacity Arena Allocator (AA)}
\label{sec:aa}

Unlike the Fixed-capacity Arena Allocator (FAA), the variable-capacity Arena Allocator (AA) does not have a fixed memory budget. Instead, it dynamically allocates additional pools as needed using \texttt{new[]} from the C++ standard library. Its pseudocode is shown in Algorithm \ref{alg:aa} and, like the FAA, it operates through three primary functions: \texttt{allocate()}, \texttt{deallocate()}, and \texttt{reset()}.

\begin{algorithm}[hbtp]
\caption{A variable-capacity Arena Allocator (AA).}
\label{alg:aa}
\begin{algorithmic}[1]
\Require{$\textsc{alloc\_size}_a$: Size of each allocation (constant)}
\Require{$\textsc{pool\_size}_a$: Size of each memory pool (constant)}

\Statex

\State \textbf{struct} $AA \langle \textsc{alloc\_size}_a, \textsc{pool\_size}_a \rangle ()$
\State \ \ $\textsc{alloc\_size} \gets \textsc{alloc\_size}_a$ \Comment{Size of each allocation} \label{alg:aa--init-const-begin}
\State \ \ $\textsc{pool\_size} \gets \textsc{pool\_size}_a$ \Comment{Size of each memory pool} \label{alg:aa--init-const-end}
\State \ \ $freed \gets \{\}$ \Comment{Freed allocations, can be reused} \label{alg:aa--init-begin}
\State \ \ $used \gets \textsc{pool\_size}$ \Comment{Bytes used in the last memory pool}
\State \ \ $pools \gets \{\}$ \Comment{Memory pools} \label{alg:aa--init-end}

\Statex

\State $\rhd$ Allocate memory, of size $\textsc{alloc\_size}$
\Function{allocate}{$ $} \textbf{of} AA
  \State $\rhd$ Allocate from freed list, if available
  \If{$freed \neq \{\}$} \ReturnInline{$freed.pop()$} \label{alg:aa--allocate-freed}
  \EndIf
  \State $\rhd$ Allocate from pool
  \If{$used < \textsc{pool\_size}$} \label{alg:aa--allocate-pool-begin}
    \State $ptr \gets pools.last() + used$
    \State $used \gets used + \textsc{alloc\_size}$
    \Return{$ptr$}
  \EndIf \label{alg:aa--allocate-pool-end}
  \State $\rhd$ Allocate a new pool
  \State $ptr \gets$ Allocate $\textsc{pool\_size}$ bytes \label{alg:aa--allocate-newpool}
  \If{$ptr \neq \phi$} \label{alg:aa--allocate-new-begin}
    \State $pools.push(ptr)$
    \State $used \gets \textsc{alloc\_size}$
    \Return{$ptr$}
  \EndIf \label{alg:aa--allocate-new-end}
  \Return{$\phi$} \label{alg:aa--allocate-fail}
\EndFunction

\Statex

\State $\rhd$ Free allocated memory
\Function{deallocate}{$ptr$} \textbf{of} FAA
  \State $freed.push(ptr)$ \label{alg:aa--deallocate-push}
\EndFunction

\Statex

\State $\rhd$ Free all allocated memory
\Function{reset}{$ $} \textbf{of} FAA
  \State $freed \gets \{\}$ \label{alg:aa--reset-freedused-begin}
  \State $used \gets 0$ \label{alg:aa--reset-freedused-end}
  \ForAll{$ptr \in pools$} \label{alg:aa--reset-pools-begin}
    \State Free memory at $ptr$
  \EndFor \label{alg:aa--reset-pools-end}
\EndFunction
\end{algorithmic}
\end{algorithm}

As with FAA, the allocator $AA$ is initialized with a predefined allocation size $\textsc{alloc\_size}_a$ and memory pool size $\textsc{pool\_size}_a$ (lines \ref{alg:faa--init-const-begin}-\ref{alg:faa--init-const-end}). Additionally, it maintains a list of reusable freed allocations $freed$, tracks the number of used bytes in the most recent pool $used$, and stores references to all allocated pools in $pools$ (lines \ref{alg:aa--init-begin}-\ref{alg:aa--init-end}). Memory allocation, handled by \texttt{allocate()}, follows a three-step process. \textbf{(1)} If freed allocations are available, they are reused (line \ref{alg:faa--allocate-freed}) to minimize new allocations. \textbf{(2)} If space remains in the last allocated pool, the next block is assigned, and the used counter updates accordingly (lines \ref{alg:faa--allocate-pool-begin}-\ref{alg:faa--allocate-pool-end}). \textbf{(3)} If neither condition holds, a new pool of size $\textsc{pool\_size}$ is allocated. If successful, it is added to $pools$, and allocation proceeds (lines \ref{alg:aa--allocate-new-begin}-\ref{alg:aa--allocate-new-end}); otherwise, the function returns $\phi$ to signal failure (line \ref{alg:aa--allocate-fail}). The \texttt{deallocate()} function, as earlier, allows memory to be freed by pushing freed pointers onto the $freed$ list for reuse (line \ref{alg:aa--deallocate-push}). To reset the allocator, \texttt{reset()} clears the $freed$ list, resets the $used$ counter, and releases all allocated $pools$ back to the standard library (lines \ref{alg:aa--reset-freedused-begin}-\ref{alg:aa--reset-pools-end}).

\subsubsection{Power-of-2 size Arena Allocator (P2AA)}
\label{sec:p2aa}

Unlike the AA allocator, which only supports allocating memory blocks of a fixed size, the Power-of-2 Arena Allocator (P2AA) allows arbitrary allocation sizes, specializing in handling memory allocations that are powers of two --- similar to a slab allocator \cite{bonwick1994slab}. The pseudocode for P2AA allocator is shown in Algorithm \ref{alg:p2aa}, and it builds upon the AA allocator by maintaining separate AA sub-allocators for each power-of-2 size, from $16$ bytes to $8192$ bytes. The $8192$-byte limit is chosen empirically for optimal performance.

\begin{algorithm}[hbtp]
\caption{A Power-of-2 size Arena Allocator (P2AA).}
\label{alg:p2aa}
\begin{algorithmic}[1]
\Require{$\textsc{pool\_size}_a$: Size of each memory pool (constant)}

\Statex

\State \textbf{struct} $P2AA \langle \textsc{pool\_size}_a \rangle ()$
\State \ \ $\textsc{pool\_size} \gets \textsc{pool\_size}_a$ \Comment{Size of each memory pool} \label{alg:p2aa--init-const}
\State \ \ $aa16 \gets AA \langle 16, \textsc{pool\_size} \rangle ()$ \Comment{For $16$-byte allocations} \label{alg:p2aa--init-begin}
\State \ \ $aa32 \gets AA \langle 32, \textsc{pool\_size} \rangle ()$ \Comment{For $32$-byte allocations}
\State \ \ $\cdots$ \Comment{For $64$ to $4096$-byte allocations}
\State \ \ $aa8192 \gets AA \langle 8192, \textsc{pool\_size} \rangle ()$ \Comment{For $8192$-byte allocations} \label{alg:p2aa--init-end}

\Statex

\State $\rhd$ Allocate memory of size that is a power of 2
\Function{allocate}{$size$} \textbf{of} P2AA
  \State \textbf{switch} $(size)$
  \State \ \ \textbf{case} $16$: \ReturnInline{$aa16.allocate()$} \label{alg:p2aa--allocate-pow2-begin}
  \State \ \ \textbf{case} $32$: \ReturnInline{$aa32.allocate()$}
  \State \ \ $\cdots$
  \State \ \ \textbf{case} $8192$: \ReturnInline{$aa8192.allocate()$} \label{alg:p2aa--allocate-pow2-end}
  \State \ \ \textbf{default}: \ReturnInline{Allocate $size$ bytes} \label{alg:p2aa--allocate-other}
\EndFunction

\Statex

\State $\rhd$ Free allocated memory
\Function{deallocate}{$ptr, size$} \textbf{of} P2AA
  \State \textbf{switch} $(size)$
  \State \ \ \textbf{case} $16$: $aa16.deallocate(ptr)$ \label{alg:p2aa--deallocate-pow2-begin}
  \State \ \ \textbf{case} $32$: $aa32.deallocate(ptr)$
  \State \ \ $\cdots$
  \State \ \ \textbf{case} $8192$: $aa8192.deallocate(ptr)$ \label{alg:p2aa--deallocate-pow2-end}
  \State \ \ \textbf{default}: Free memory at $ptr$ \label{alg:p2aa--deallocate-other}
\EndFunction

\Statex

\State $\rhd$ Free all allocated memory
\Function{reset}{$ $} \textbf{of} P2AA
  \State $aa16.reset()$ \label{alg:p2aa--reset-begin}
  \State $aa32.reset()$
  \State $\cdots$
  \State $aa8192.reset()$ \label{alg:p2aa--reset-end}
\EndFunction

\Statex

\State $\rhd$ Get recommended allocation size, for a desired size
\Function{allocationSize}{$size$} \textbf{of} P2AA
  \If{$size \leq 16$} \ReturnInline{$16$} \label{alg:p2aa--allocationsize-16}
  \ElsIf{$size < 8192$} \ReturnInline{$nextPow2(size)$} \label{alg:p2aa--allocationsize-8192}
  \Else\ \ReturnInline{$\lceil size / \textsc{page\_size} \rceil * \textsc{page\_size}$} \label{alg:p2aa--allocationsize-other}
  \EndIf
\EndFunction
\end{algorithmic}
\end{algorithm}

At initialization, the allocator $P2AA$ is configured with a predefined memory pool size $\textsc{pool\_size}_a$ (line \ref{alg:p2aa--init-const}). It creates multiple AA sub-allocators, each responsible for a different power-of-2 allocation size, ranging from $16$ to $8192$ bytes (lines \ref{alg:p2aa--init-begin}-\ref{alg:p2aa--init-end}). Memory allocation is performed in the \texttt{allocate()} function. When a request for a standard power-of-2 size is made (e.g., $16$, $32$, or $8192$ bytes), the request is routed to corresponding AA sub-allocator handles the request (lines \ref{alg:p2aa--allocate-pow2-begin}-\ref{alg:p2aa--allocate-pow2-end}). For sizes larger than $8192$ bytes or non-power-of-2 sizes, a direct memory allocation is performed instead (line \ref{alg:p2aa--allocate-other}). Deallocation mirrors the allocation strategy. Each deallocation request, made through \texttt{deallocate()}, is dispatched to the corresponding sub-allocator, if it is a power-of-two, and ranges from $16$ to $8192$ bytes in size (lines \ref{alg:p2aa--deallocate-pow2-begin}-\ref{alg:p2aa--deallocate-pow2-end}). For other sizes, memory is freed directly using \texttt{delete[]} (line \ref{alg:p2aa--deallocate-other}). Note that, unlike \texttt{delete[]}, the user is expected to provide the size of the allocated memory. To reset all allocated memory, the \texttt{reset()} function (lines \ref{alg:p2aa--reset-begin}-\ref{alg:p2aa--reset-end}) resets each individual arena allocator. This effectively clears all memory pools, making the allocator ready for fresh allocations without explicitly deallocating each individual allocation. Finally, the \texttt{allocationSize()} function provides a mechanism for determining the optimal allocation size for a given request. If the request is $16$ bytes or smaller, it returns $16$ (line \ref{alg:p2aa--allocationsize-16}). If the request is between $17$ and $8191$ bytes, it rounds up to the next power of two using \texttt{nextPow2()} (line \ref{alg:p2aa--allocationsize-8192}). For larger requests, it rounds up to the nearest multiple of the system page size (line \ref{alg:p2aa--allocationsize-other}). This ensures that large memory allocations are page-aligned.

\subsubsection{Concurrent Power-of-2 Arena Allocator (CP2AA)}
\label{sec:cp2aa}

We now discuss our thread-safe Concurrent Power-of-2 Arena Allocator (CP2AA), which builds upon per-thread instances of the Power-of-2 Arena Allocator (P2AA). Its pseudocode is presented in Algorithm \ref{alg:cp2aa}. As above, CP2AA consists of four primary functions: \texttt{allocate()}, \texttt{deallocate()}, \texttt{reset()}, and \texttt{allocationSize()}.

\begin{algorithm}[hbtp]
\caption{Our Concurrent Pow-of-2 Arena Allocator (CP2AA).}
\label{alg:cp2aa}
\begin{algorithmic}[1]
\Require{$\textsc{pool\_size}_a$: Size of each memory pool (constant)}

\Statex

\State \textbf{struct} $CP2AA \langle \textsc{pool\_size}_a \rangle ()$
\State \ \ $\textsc{pool\_size} \gets \textsc{pool\_size}_a$ \Comment{Size of each memory pool} \label{alg:cp2aa--init-const}
\State \ \ $p2aa_T \gets \{P2AA \langle \textsc{pool\_size} \rangle (),\ \dots\}$ \Comment{Per-thread allocator} \label{alg:cp2aa--init}

\Statex

\State $\rhd$ Allocate memory of size that is a power of 2
\Function{allocate}{$size$} \textbf{of} CP2AA
  \State $t \gets$ Current thread \label{alg:cp2aa--allocate-begin}
  \Return{$p2aa_T[t].allocate(size)$} \label{alg:cp2aa--allocate-end}
\EndFunction

\Statex

\State $\rhd$ Free allocated memory
\Function{deallocate}{$ptr, size$} \textbf{of} CP2AA
  \State $t \gets$ Current thread \label{alg:cp2aa--deallocate-begin}
  \State $p2aa_T[t].deallocate(ptr, size)$ \label{alg:cp2aa--deallocate-end}
\EndFunction

\Statex

\State $\rhd$ Free all allocated memory
\Function{reset}{$ $} \textbf{of} CP2AA
  \ForAll{$t \in threads$} \label{alg:cp2aa--reset-begin}
    \State $p2aa_T[t].reset()$
  \EndFor \label{alg:cp2aa--reset-end}
\EndFunction

\Statex

\State $\rhd$ Get recommended allocation size, for a desired size
\Function{allocationSize}{$size$} \textbf{of} CP2AA
  \Return{$P2AA \langle \textsc{pool\_size} \rangle.allocationSize(size)$} \label{alg:cp2aa--allocationsize}
\EndFunction
\end{algorithmic}
\end{algorithm}


The $CP2AA$ allocator is initialized with a fixed memory pool size $\textsc{pool\_size}_a$ (line \ref{alg:cp2aa--init-const}), and maintains a separate instance of P2AA per thread, stored in $p2aa_T$ (line \ref{alg:cp2aa--init}) --- while ensuring that each P2AA allocator is well-separated in memory to prevent false sharing. Each thread exclusively interacts with its corresponding P2AA instance, eliminating the need for locks or atomic operations altogether. The \texttt{allocate()} function assigns memory in a thread-local manner. Given a requested size, it identifies the current thread $t$ and delegates the allocation to its corresponding P2AA instance, $p2aa_T[t]$ (lines \ref{alg:cp2aa--allocate-begin}-\ref{alg:cp2aa--allocate-end}). Similarly, \texttt{deallocate()} retrieves the current thread $t$ and forwards the deallocation request to the P2AA allocator of the current thread (lines \ref{alg:cp2aa--deallocate-begin}-\ref{alg:cp2aa--deallocate-end}). Note that it is acceptable for a thread to deallocate memory allocated by another thread. In fact, this lack of restriction is a key source of performance improvement. The \texttt{reset()} function deallocates all of the allocated memory by iterating over all per-thread allocators and calling their respective reset functions (lines \ref{alg:cp2aa--reset-begin}-\ref{alg:cp2aa--reset-end}). Finally, the \texttt{allocationSize()} function determines the appropriate power-of-2 allocation size for a given request by forwarding the query to P2AA (line \ref{alg:cp2aa--allocationsize}).

\ignore{Our initial attempt at a concurrent arena allocator used an atomic\_flag mutex to manage freed blocks, but high contention resulted in slow deallocation, especially with 64 threads.}

\subsubsection{Performance Comparison}

\begin{figure}[hbtp]
  \centering
  \subfigure[Allocation-only workload: $2^{28}$ allocations of 64 bytes.]{
    \label{fig:allocator-alloc--runtime}
    \includegraphics[width=0.98\linewidth]{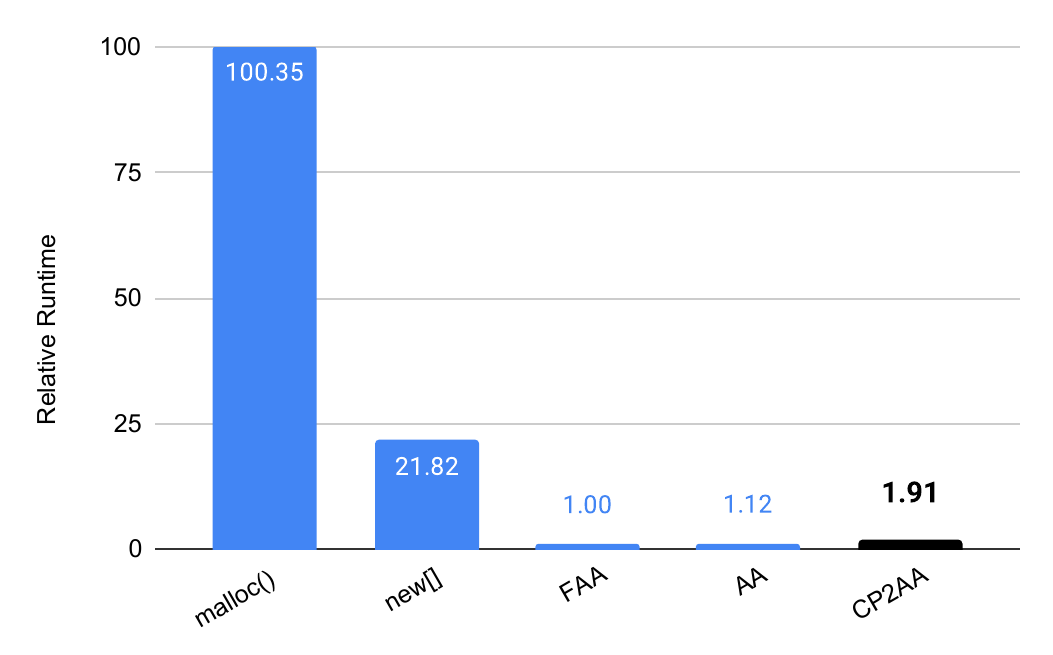}
  }
  \subfigure[Deallocation-only workload: $2^{28}$ deallocations of allocated memory.]{
    \label{fig:allocator-free--runtime}
    \includegraphics[width=0.98\linewidth]{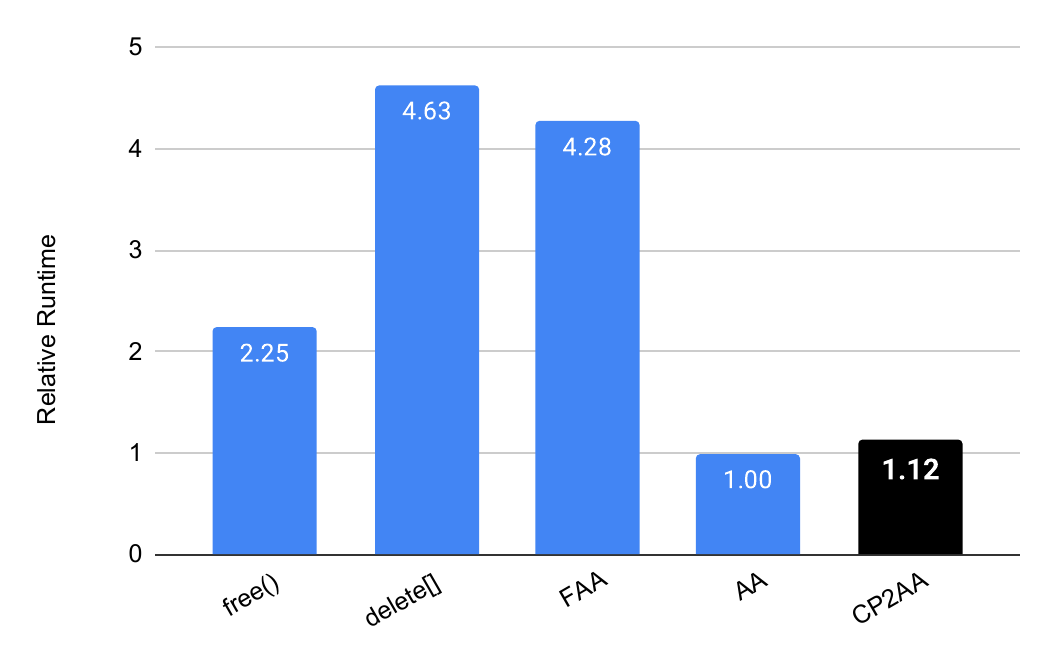}
  }
  \subfigure[Mixed workload: $2^{22}$ allocations and $2^{22}$ deallocations, $64$ times.]{
    \label{fig:allocator-mixed--runtime}
    \includegraphics[width=0.98\linewidth]{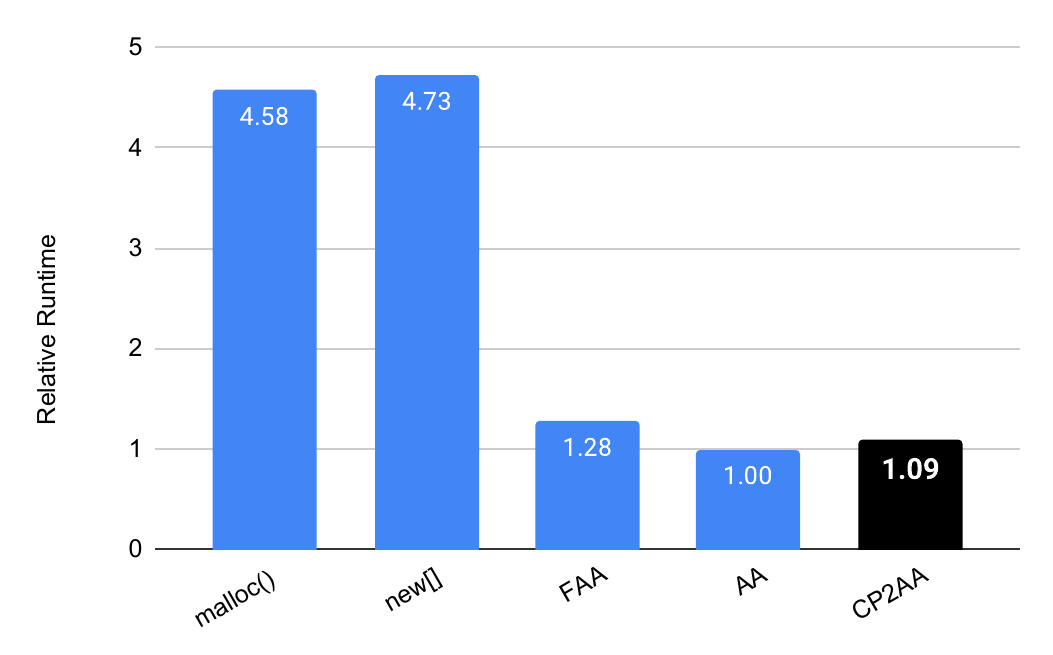}
  } \\[-2ex]
  \caption{Relative Runtime of memory allocators across three workloads: \textit{(a)} \textit{allocation-only}, where $2^{28}$ allocations of $64$ bytes each are performed; \textit{(b)} \textit{deallocation-only}, with $2^{28}$ deallocations; and \textit{(c)} mixed, with $2^{22}$ allocations followed by $2^{22}$ deallocations, repeated $64$ times. The allocators include the C library allocator ($malloc()/free()$), C++ runtime allocator ($new[]/delete[]$), Fixed Arena Allocator (FAA), variable-capacity Arena Allocator (AA), and Concurrent Power-of-2 Arena Allocator (CP2AA). Details on FAA, AA, and CP2AA allocators is given in Section \ref{sec:memory-allocators}.}
  \label{fig:allocator-runtime}
\end{figure}

To evaluate the performance of different memory allocators, we conduct three separate experiments, each designed to measure execution time under specific memory management workloads. The allocators tested include the C library allocator (\texttt{malloc()}/\texttt{free()}), the C++ runtime allocator (\texttt{new[]}/\texttt{delete[]}), the Fixed Arena Allocator (FAA), the variable-capacity Arena Allocator (AA), and the Concurrent Power-of-2 Arena Allocator (CP2AA). In the first experiment, we measure the performance of each allocator in an allocation-heavy scenario. Here, each allocator is subject to a total of $2^{28}$ memory allocations, each of size $64$ bytes. With non-concurrent allocators, i.e., FAA and AA, separate allocator objects are created for each thread to ensure thread safety, and the workload is distributed evenly across all threads. The allocated memory addresses are stored in an array to ensure they can be deallocated in the subsequent experiment. In the second experiment, we evaluate the deallocation performance of each allocator. Using the memory addresses obtained from the first experiment, each allocator is tasked with deallocating $2^{28}$ blocks of memory. Deallocations are routed to the appropriate allocator object for each thread in the case of FAA and AA. The third experiment examines allocator performance in a mixed workload, where allocation and deallocation operations occur in succession. Specifically, each allocator is tested with $2^{22}$ allocations followed by $2^{22}$ deallocations, with this cycle being repeated $64$ times. This experiment provides insight into the performance of each allocator under workloads that mimic real-world application behavior.

Figure \ref{fig:allocator-alloc--runtime} presents the relative runtime of each allocator for the allocation-only workload, while Figure \ref{fig:allocator-free--runtime} shows the deallocation-only workload. Finally, Figure \ref{fig:allocator-mixed--runtime} illustrates the performance of each allocator under the mixed workload. The results indicate that specialized allocators like FAA, AA, and CP2AA offer substantial performance improvements over general-purpose allocators (\texttt{malloc()}/\texttt{new[]}), particularly in allocation-intensive workloads, and offer around $4\times$ speedup in mixed workloads. While the AA allocator performs the best, on average, it is not thread-safe and thus not suitable for concurrent applications. The CP2AA allocator, which is both thread-safe and high-performing, is suitable.

\subsection{Evaluating Graph Representations via Graph Algorithm Performance}

As dicusssed earlier, we evaluate the performance of various graph representations on a representative algorithm by measuring the efficiency of $k$-step reverse walks from each vertex in the given input graph, and counting the number of walks ending at each vertex. This corresponds to computing $A_T^k \cdot \vec{1}$, where $A_T$ is the transposed adjacency matrix and $\vec{1}$ is a ones vector. Reverse walks are preferred as they can be executed directly on the input graph, whereas forward walks require its transpose. The results of this evaluation on each graph representation, i.e., PetGraph, SuiteSparse:GraphBLAS, cuGraph, Aspen, and our DiGraph, are discussed in Section \ref{sec:perform-reverse-walks}.

We now give a short description of the algorithm. It propagates visit counts backward along graph edges, and its psuedocode is given in Algorithm \ref{alg:visit}. Here, the \texttt{reverseWalk()} function, takes a graph $G$ and the number of reverse walk $steps$ as input. Initially, two arrays, $visits0$ and $visits1$, are allocated to track visit counts, with $visits0$ set to $1$ for all vertices and $visits1$ initialized to $0$ (lines \ref{alg:visit--init-begin}-\ref{alg:visit--init-end}). The main loop iterates $steps$ times, processing all vertices $u$ in parallel. Each iteration resets $visits1[u]$ (line \ref{alg:visit--reset}) and accumulates the visit count from each neighbor $v$ of $u$ (lines \ref{alg:visit--edges-begin}-\ref{alg:visit--edges-end}). This iterative process propagates visit counts backward through the graph. At the end of each iteration, the visit count arrays are swapped (line \ref{alg:visit--swap}), ensuring $visits0$ contains the latest counts. After the final iteration, $visits0$ contains the number of reverse walks ending at each vertex\ignore{in the graph, and is returned}.

\begin{algorithm}[hbtp]
\caption{Perform reverse walk from each vertex.}
\label{alg:visit}
\begin{algorithmic}[1]
\Require{$G(V, E)$: Input graph}
\Ensure{$steps$: Number of reverse walks to perform}

\Statex

\Function{reverseWalk}{$G, steps$}
  \State $visits0 \gets \{1\}$ \label{alg:visit--init-begin}
  \State $visits1 \gets \{0\}$ \label{alg:visit--init-end}
  \ForAll{$i \in [0, steps)$} \label{alg:visit--steps-begin}
    \ForAll{$u \in V$ \textbf{in parallel}}
      \State $visits1[u] \gets 0$ \label{alg:visit--reset}
      \ForAll{$(v, \_) \in G.edges(u)$} \label{alg:visit--edges-begin}
        \State $visits[u] \gets visits[u] + visits0[v]$
      \EndFor \label{alg:visit--edges-end}
    \EndFor
    \State $swap(visits0, visits1)$ \label{alg:visit--swap}
  \EndFor \label{alg:visit--steps-end}
  \Return{$visits0$}
\EndFunction
\end{algorithmic}
\end{algorithm}

\end{document}